\documentclass[preprint,journal]{arxiv}            % preprint (journal style) arxiv no copyright

\onlineid{1546}

\vgtccategory{Research}

\vgtcpapertype{Technique}

\title{HyperINR: A Fast and Predictive Hypernetwork for Implicit Neural Representations via Knowledge Distillation}

\author{%
  \authororcid{Qi~Wu}{0000-0002-1825-0097},
  \authororcid{David~Bauer}{0000-0000-0000-0000},
  \authororcid{Yuyang~Chen}{0000-0000-0000-0000},
  and~\authororcid{Kwan-Liu~Ma}{0000-0000-0000-0000}
}

\authorfooter{
  \item Authors are with the University of California, Davis.
  E-mail: {\normalfont\{}qadwu\,$|$\,davbauer\,$|$\,yuach\,$|$\,klma{\normalfont\}}@ucdavis.edu\,.
}

\abstract{% % 
Implicit Neural Representations (INRs) have recently exhibited immense potential in the field of scientific visualization for both data generation and visualization tasks. However, these representations often consist of large multi-layer perceptrons (MLPs), necessitating millions of operations for a single forward pass, consequently hindering interactive visual exploration. While reducing the size of the MLPs and employing efficient parametric encoding schemes can alleviate this issue, it compromises generalizability for unseen parameters, rendering it unsuitable for tasks such as temporal super-resolution.
In this paper, we introduce HyperINR, a novel hypernetwork architecture capable of directly predicting the weights for a compact INR. By harnessing an ensemble of multiresolution hash encoding units in unison, the resulting INR attains state-of-the-art inference performance (up to $100\times$ higher inference bandwidth) and supports interactive photo-realistic volume visualization. Additionally, by incorporating knowledge distillation, exceptional data and visualization generation quality is achieved, making our method valuable for real-time parameter exploration. We validate the effectiveness of our HyperINR architecture through a comprehensive ablation study.
We showcase the versatility of HyperINR across three distinct visualization tasks: novel view synthesis, temporal super-resolution of volume data, and volume rendering with dynamic global shadows. By simultaneously achieving efficiency and generalizability, HyperINR paves the way for applying INR in a wider array of scientific visualization applications.
}


\teaser{
  \centering
  \includegraphics[width=\linewidth]{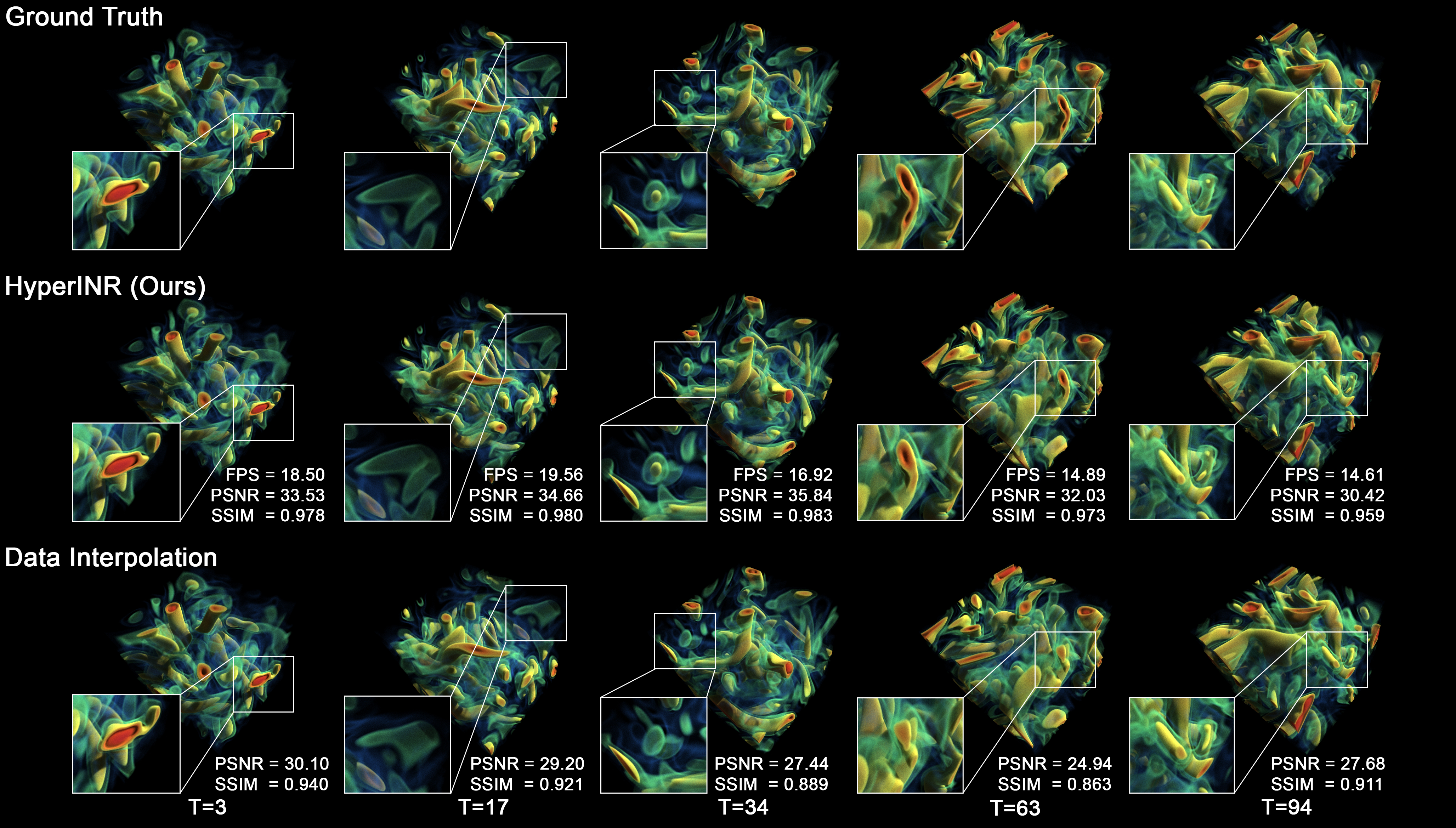}
  \vspace{-1.8em}
  \caption{Comparisons between HyperINR and data interpolation for the temporal super-resolution task using the \textbf{vortices} dataset. Listed timesteps are midpoints of different interpolation intervals. HyperINR can directly predict the weights of a regular implicit neural representation (INR) for unseen parameters. The predicted INR is in general more accurate than data interpolation results and can support interactive volumetric path tracing.}
  \label{fig:teaser}
}

\graphicspath{{figs/}{figures/}{pictures/}{images/}{./}} % where to search for the images

\usepackage{tabu}                      % only used for the table example
\usepackage{booktabs}                  % only used for the table example
\usepackage{lipsum}                    % used to generate placeholder text
\usepackage{mwe}                       % used to generate placeholder figures

\usepackage{mathptmx}                  % use matching math font

\usepackage[dvipsnames,svgnames]{xcolor}
\usepackage{xspace}
\usepackage{soul}
\usepackage{comment}
\usepackage{multirow}
\usepackage{listings}
\usepackage{amsmath}
\usepackage{amssymb}
\usepackage{adjustbox}
\usepackage{enumitem}
\usepackage[normalem]{ulem}
\usepackage{multicol}
\usepackage{makecell}
\usepackage{tabu}
\usepackage{booktabs, siunitx}
\usepackage{algpseudocode}
\usepackage{algorithm}
\usepackage[mathcal]{euscript}
\usepackage[utf8]{inputenc}
\usepackage[T1]{fontenc}

\definecolor{codeGreen}{rgb}{0,0.6,0}
\definecolor{codeBlue}{rgb}{0,0,1}
\definecolor{codeRed}{rgb}{0.65,0.11,0.36}
\definecolor{codeGray}{rgb}{0.6,0.6,0.6}
\definecolor{codeMauve}{rgb}{0.58,0,0.82}
\definecolor{codeCyan}{rgb}{0,0.52,0.70}

\newcommand{\ie}{\MakeLowercase{i.e.,}\xspace}

\newcommand{\etal}{\MakeLowercase{et al.}\xspace}

\begin{document}

%%%%%%%%%%%%%%%%%%%%%%%%%%%%%%%%%%%%%%%%%%%%%%%%%%%%%%%%%%%%%%%%
%%%%%%%%%%%%%%%%%%%%%% START OF THE PAPER %%%%%%%%%%%%%%%%%%%%%%
%%%%%%%%%%%%%%%%%%%%%%%%%%%%%%%%%%%%%%%%%%%%%%%%%%%%%%%%%%%%%%%%

\firstsection{Introduction}

\maketitle

In the field of scientific visualization, continuous fields are often represented using discrete data structures such as grids, unstructured meshes, or point clouds. These structures are limited by resolution and can be cumbersome to handle due to their complexity. To address this, Lu~\etal~\cite{lu2021compressive} introduced an alternative approach, employing a continuous function modeled using a fully connected multi-layer perceptron (MLP) to implicitly represent data fields. Such an implicit neural representation (INR) offers several key advantages, including substantial data size reductions while preserving high-frequency details, and direct access to spatial locations at arbitrary resolutions without decompression or interpolation.
% \cad{... This INR offers several key advantages, including substantial data size reductions while preserving high-frequency details, and direct access to spacial locations at arbitrary resolutions without decompression or interpolation.}
% \cad{One of such advancements in INR would be CoordNet, introduced by Han, which generalizes...}
% 
A recent advancement in INR is CoordNet, introduced by Han~\etal~\cite{han2022coordnet}, which generalizes INR to incorporate simulation parameters  ($\theta_{sim}$) and visualization parameters ($\theta_{vis}$) for predictive tasks such as temporal super-resolution or visualization synthesis. With a well-designed fully connected INR and a suitable activation function, CoordNet generates more meaningful results than direct data interpolation for previously unseen parameters. However, a single pass through such a large fully connected INR may require millions of operations, making it slow for neural network inference and thus unsuitable for interactive visualizations.

% \cad{I feel like a transition might be needed here to shift the topic from CoordNet's long inference time to your work in reducing network size. There might be a confusion where people relate the previous paragraph's 'coordNet' to the 'neural network' that you talk about below...nevermind, fixed it. see below.}

Wu~\etal~\cite{wu2022instant} and Weiss~\etal~\cite{weiss2021fast} addressed the challenge of long inference times of large fully connected INRs by reducing the neural network size to approximately 200 neurons and incorporating additional trainable parameters through auxiliary data structures to compensate for the reduction in network capacity. Such auxiliary data structures are typically constructed in the form of hash tables~\cite{muller2022instant} or octrees~\cite{takikawa2021neural}, and are responsible for transforming input coordinates into high-dimensional vectors, a process known as parametric positional encoding. Compared with parameters stored in the neural network, querying, interpolating, and optimizing encoding parameters based on input coordinates can be performed more rapidly. Consequently, interactive visualization of INR models can be achieved. 
% 
% However, due to the majority of data features being locally over-fitted by encoding parameters rather than learned by the fully connected MLP, the generalizability of the INR is significantly reduced. 
% 
However, the majority of data features are strongly embedded into the trainable encoding parameters rather than being generalized by the MLP. In turn, the generalizability of INR in regards to unseen data is significantly reduced.
Therefore, a method that can offer strong generalizability while still maintaining exceptional inference performance for enabling interactive visualization and real-time parameter exploration is highly desirable.

In this paper, we present HyperINR, a hypernetwork designed to conditionally predict the weights of an Implicit Neural Representation (INR) using multiple compact multiresolution hash encoders~\cite{muller2022instant} and incorporating a deeply embedded weight interpolation operation. These hash encoders correspond to points in the parameter space and are organized within a spatial data structure.
Given an input, the data structure is traversed to gather a set of nearest hash encoders, whose weights are subsequently interpolated based on the input and combined with the weights of a shared MLP common to all encoders. This combined weight is then suitable for use as an INR, enabling interactive volumetric path tracing through the state-of-the-art INR rendering algorithm proposed by Wu~\etal~\cite{wu2022instant}, with the INR generation process taking less than 1\textit{ms} to complete (up to $100\times$ faster than CoordNet). Furthermore, we optimize HyperINR using knowledge distillation, leveraging CoordNet~\cite{han2022coordnet} as the teacher model, to achieve state-of-the-art generalizability for unseen parameters, rendering HyperINR suitable for real-time parameter exploration. We conduct a comprehensive evaluation of HyperINR's architecture through an extensive ablation study and assess its performance in three distinct scientific visualization tasks: Novel View Synthesis (NVS), Temporal Super-Resolution of Volume Data (TSR), and Volume Rendering with Dynamic Global Shadows (DGS).
Our contributions can be summarized as follows. 
\begin{itemize}
    \item We design HyperINR: A hypernetwork that efficiently generates the weights of a regular INR for given parameters, achieving state-of-the-art inference performance and enabling high-quality interactive volumetric path tracing.
    \item We introduce a framework for optimizing HyperINR through knowledge distillation, attaining state-of-the-art data generalization quality for unseen parameters and supporting real-time parameter exploration.
    \item We demonstrate HyperINR's exceptional inference performance and its ability to generate meaningful data and visualizations across a diverse range of scientific visualization tasks.
\end{itemize}

\begin{figure*}[tbp]
  \centering
  \includegraphics[width=0.9\textwidth]{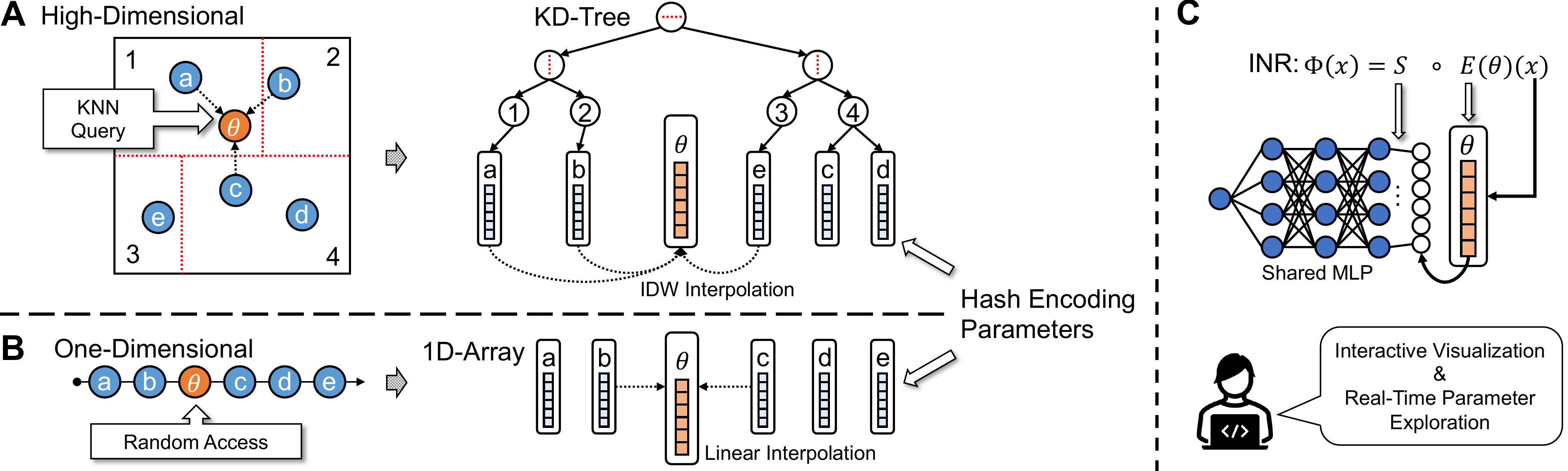}
  \vspace{-0.8em}
  \caption{The architecture of HyperINR. HyperINR is composed of a shared MLP denoted as $S$ and a collection of multiresolution hash encoding parameters $E$, with each parameter associated with a set of values $\{a, b, c, ...\}$. A) In high dimensions, a KD-tree is utilized, B) while in 1D, the encoding parameters are arranged in a linear array. C) The input parameter $\theta$ is utilized to query the data structure, and the queried encoding parameters are interpolated and combined with the shared MLP to construct an INR. The resulting INR enables interactive visualization and real-time parameter exploration.}
  \label{fig:overview}
  \vspace{-1em}
\end{figure*}

%%%%%%

% \qwu{commented comments temporarily to see the page limit. thanks.}

% \qwu{Consider highlighting the connection with volume compression.}
% \qwu{Is generalizability a good choice of word? In title I used predictive. What I really want to say is the ability to generate meaningful data for unseen parameters.}

% \cad{"generalizability in denotation does not fit this well, but somewhat works in connotation (albeit still inaccurate). predictive works in denotation, but not many people understand it connotation-wise"}

% \cad{I am probably worse in English than you are. }

% \cad{A highlight for HyperINR because I had to read a couple times to get what it does: HyperINR is a hypernetwork method to generate weights for a regular INR by feeding the input parameter through a KD-Tree worth of HashGrids, KNN sample that tree, LERP, and concat with an MLP}

% \cad{Second bullet point nitpicking the grammar: "the employment of.." feels inconsistent to the list of "contribution". We would expect a simple noun object with the rest of list starting with "HyperINR", and "a demonstration". I would phrase it as: "A method to optimize HyperINR with the employment of knowledge distillation, ..." for more consistency.}

\section{Related Work}

In this related work section, we delve into the relevant research areas related to our presented work. We begin by providing an overview of recent advancements in generation models for scientific visualization, 
followed by a review of implicit neural representation. 
Because our technique involves hypernetwork and knowledge distillation, we also review the latest advancements in these areas.

\begin{comment}

\begin{table}[tb]
  \caption{Notations involved in this paper.}
  \label{tab:notations}
  \scriptsize%
  \centering%
  \begin{tabu}{cl}
    \toprule
    Notation & Meaning \\
    \midrule
    $\Phi$    & The function parameterized by a regular INR \\
    $H$       & The higher-order function parameterized by HyperINR \\
    $E$       & The function representing a multiresolution hash encoding  \\
    $S$       & The function representing an MLP \\
    \midrule
    $V$       & An image or volume data \\
    $\Theta$  & A set of scene parameters, inputs for HyperINR \\
    \midrule
    $N_E$     & The number of multiresolution hash encoding units \\
    $N_D$     & The number of parameters sampled for knowledge distillation \\
    $K$       & The size of a K-nearest neighborhood query \\
    $L$       & The number of encoding levels \\
    $T$       & The multiresolution hash encoding table size \\
    $F$       & The number of features per hash encoding level \\
    \bottomrule
  \end{tabu}%
\end{table}

\end{comment}

\subsection{Generation Models for Scientific Visualization}\label{sec:related-generative-model}

Using neural networks to generate new data or visualization that are similar to the training data has been studied by many in recent years.
We overview  related in this area, and we refer to Wang~\etal~\cite{wang2022dl4scivis} for a more comprehensive survey of this topic.

For visualization generation,
Berger~\etal~\cite{berger2018generative} developed a generative adversarial network (GAN) for synthesizing volume rendering images with different transfer functions and view parameters.
GAN was also used by He~\etal~\cite{he2019insitunet} to create a simulation and visualization surrogate model called InSituNet for exploring ensemble simulation parameters.
Engel and Ropinski~\cite{engel2020deep} built a 3D U-Net that can predict local ambient occlusion data for given transfer functions. 
Weiss~\etal~\cite{weiss2019volumetric} proposed a convolutional neural network (CNN) network to coherently upscale isosurface images by training the network using depth
and normal information. 
Han and Wang~\cite{han2022vcnet} developed a GAN-based volume completion network for visualizing data with missing subregions.
Bauer~\etal~\cite{bauer2023fovolnet} introduced a CNN-based screen space method that enables faster volume rendering through sparse sampling and neural reconstruction.

For data generation, 
Zhou~\etal~\cite{zhou2017volume} and Guo~\cite{guo2020ssr} designed CNNs for upscaling scalar and vector field volume data respectively.
Han and Wang employed GANs for generating time-varying volume data at higher temporal~\cite{han2019tsr} or spatial~\etal~\cite{han2020ssr} resolutions. 
Han~\etal~\cite{han2021stnet} later also presented an end-to-end solution for achieving both goals at the same time.
Shi~\etal~\cite{shi2022vdl} improved InSituNet through view-dependent latent-space generative models, and their method can directly predict simulation data rather than being bound to the visualization strategy used for generating the training data.
Data generation models are also applicable to volume compression.
Jain~\etal~\cite{jain2017compressed} presented an encoder-decoder network to compress a high-resolution volume.
Wurster~\etal~\cite{wurster2021deep} developed a hierarchical GAN for the same task. 
 
Very recently, Han~\etal~\cite{han2022coordnet} explored the use of implicit neural representation and proposed CoordNet for both data and visualization generation tasks. In our work, we use CoordNet as the teacher model for knowledge distillation.

\subsection{Implicit Neural Representation (INR)}

In the field of scientific visualization, Lu~\etal~\cite{lu2021compressive} first explored the use of INR for volume compression. 
Their network utilizes multiple ResNet blocks and the sine activation function to achieve high reconstruction qualities.
Previously mentioned CoordNet by Han~\etal~\cite{han2022coordnet} for time-varying data is also worth mentioning here. CoordNet can be considered as a conditional-INR.
However, their method requires a time-consuming training process for every volume data, and more crucially, the network is slow for inference.
Wu~\etal~\cite{wu2022instant} and Weiss~\etal~\cite{weiss2021fast} addressed these issues using parametric positional encoding~\cite{takikawa2021neural, muller2022instant} and GPU-accelerated inference routines.
Wu~\etal also proposed an auxiliary data structure to enable interactive volume rendering and an optimized algorithm for computing global illuminations.
In this work, we adopt the INR architecture proposed by Wu~\etal as the base. We develop our interactive volume visualization algorithms based on their proposals.
INR can also be used in hybrid with other hierarchical data structures. Doyub~\etal~\cite{kim2022neuralvdb} recently demonstrated this for handling high-resolution sparse volumes.

Positional encoding converts an input coordinate to a higher-dimension vector before being passed to subsequent layers. It allows the network to capture high-frequency local details better. 
Positional encoding was proven to be helpful in the attention components of recurrent networks~\cite{gehring2017convolutional} and transformers~\cite{vaswani2017attention}, and later adopted by NeRF~\cite{mildenhall2020nerf} and many INR-based works~\cite{tancik2020fourier, barron2021mip, muller2019neural} in computer graphics.
To further optimize training time and improve accuracy, \emph{parametric positional encoding} was introduced. It introduces an auxiliary  data structures such as dense grids~\cite{martel2021acorn}, sparse grids~\cite{hadadan2021neural}, octrees~\cite{takikawa2021neural}, or multiresolution hash tables~\cite{muller2022instant} to store training parameters. 
Thus, the neural network size can be reduced. 
Therefore neural networks with such encoding methods can typically converge much faster. 
In this work, we adopt the multiresolution hash grid method proposed by M{\"u}ller~\etal~\cite{muller2022instant} due to its excellent performance.

\subsection{Hypernetwork for INR}

Hypernetworks or meta-models  are  networks that generate weights for other neural networks~\cite{ha2016hypernetworks}. They have a wide range of applications, including few-shot learning~\cite{bertinetto2016learning}, continual learning~\cite{von2019continual}, architecture search~\cite{zhang2018graph}, and generative modeling~\cite{ratzlaff2019hypergan, nguyen2020hypervae, oh2020hcnaf}, among others.
Hypernetworks can also be combined with implicit neural representations (INRs). For instance, Klocek~\etal~\cite{klocek2019hypernetwork} developed an INR-based hypernetwork for image super-resolution. DeepMeta~\cite{littwin2019deep} builds a hypernet that takes a single-view image and outputs an INR. Skorokhodov~\etal~\cite{skorokhodov2021adversarial} developed a GAN-based hypernet for continuous image generation. Sitzmann~\etal~\cite{sitzmann2019scene} proposed an MLP-based hypernetwork to parameterize INRs for 3D scenes consisting of only opaque surfaces.
In this work, we utilize hypernetworks to build a large neural network that has the potential to learn an ensemble of data while maintaining the capability of interactive volume visualization.
% 
% Another important line of research is conditional INRs~\cite{}. \qwu{add missing information later.}
% \qwu{Highlight again how hypernet is different from conditional-INR? Or maybe this is obvious.}

\section{Formulation for HyperINR}

INRs can be regarded as functions mapping 2D or 3D coordinates $\vec{x}$ to their corresponding field values $\vec{v}$:
\begin{equation}\label{eq:inr}
    \Phi: \vec{x} \mapsto \Phi(\vec{x}) = \vec{v},~\vec{x}\in \mathbb{R}^{2}~\text{or}~ \mathbb{R}^{3}.
\end{equation}
We employ $\Phi$ for approximating image or volume data $V$ that are intrinsically parameterized by a scene parameter (e.g., timestep, lighting direction) from a high-dimensional parameter space $\Theta$. 
Our objective is to construct a neural network capable of continuously generating such an INR based on parameters in $\Theta$, using a sparsely sampled training set $\mathcal{C} = \{V(\theta_i),~\theta_i\}$. The resulting INR is expected to be suitable for interactive visualization and real-time parameter exploration.
Formally, such a neural network can be defined as a higher-order function, $S$, which accepts scene parameters $\theta$ as inputs, and yields $\Phi$ conditioned on these parameters:
\begin{equation}
    H(\theta)(\vec{x}) = \Phi(\vec{x}~|~\theta) \in V(\theta),~\text{for}~\theta\in\Theta,~\forall~\vec{x}\in\mathbb{R}^2~\text{or}~\mathbb{R}^3.
\end{equation}
To incorporate the state-of-the-art rendering algorithm by Wu~\etal~\cite{wu2022instant} for interactive 3D visualization of INRs, we further decompose $\Phi$ into two distinct functions: an encoding function $E$, mapping input coordinates to high-dimensional vectors via a multiresolution hash encoder; and a synthesis function $S$, converting the high-dimensional vectors into data values parameterized by an MLP:
\begin{equation}
    \Phi(\vec{x}~|~\theta) = S \circ E(\theta) (\vec{x}).
\end{equation}

\begin{figure}[tbp]
  \centering
  \includegraphics[width=\linewidth]{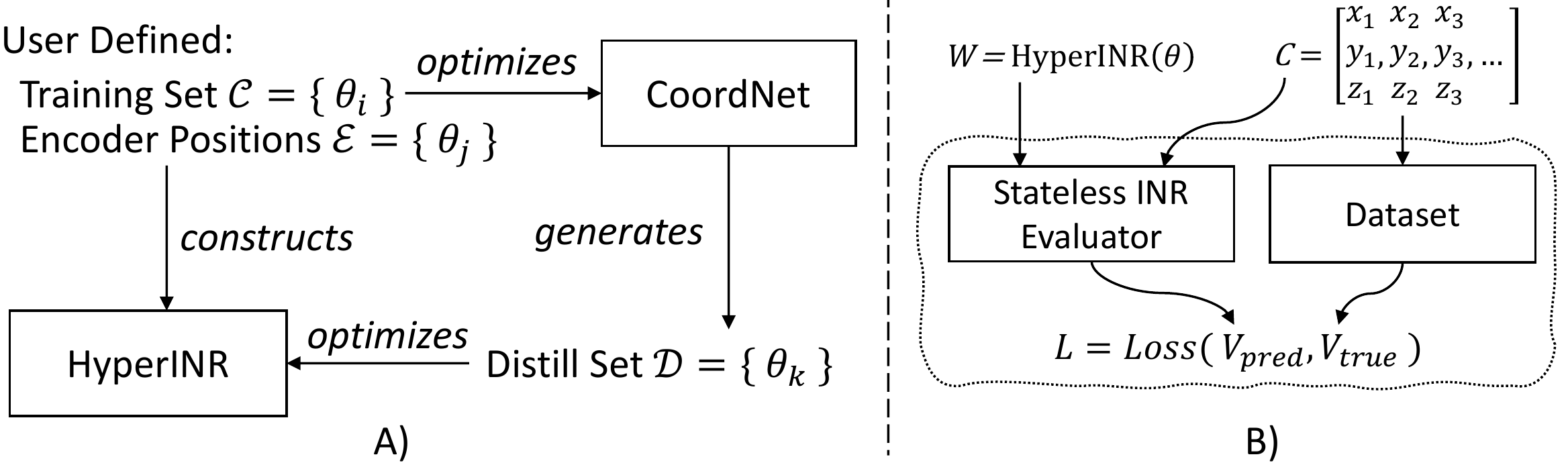}
  \vspace{-2em}
  \caption{A) Visualization of the knowledge distillation process, wherein a user provides a training set $\mathcal{C}$ and a set of encoder positions $\mathcal{E}$. We utilize $\mathcal{C}$ to pretrain CoordNet and construct HyperINR using $\mathcal{E}$. Following pretraining, we create a distillation set $\mathcal{D}$ and optimize HyperINR using it. B) We introduce a stateless INR evaluator for end-to-end training of the hypernetwork, leading to improved training quality.}
  \label{fig:eval}
  \vspace{-1em}
\end{figure}

In our work, we utilize a shared synthesis function $S$ for all scene parameters, intentionally designed to significantly enhance training robustness, as demonstrated through our experimental results in \Cref{sec:ablation-shared-mlp}.
Thus, the primary objective of HyperINR becomes predicting the encoding function $E$ for a given set of parameters $\theta$.
As illustrated in \Cref{fig:overview}, we first sample $N$ scene parameters $\mathcal{E} = \{ \theta_j~|~j=a, b, c, ... \}$ and construct a multiresolution hash encoder $E_j = E(\theta_j)$ for each $\theta_j$. We refer to $\mathcal{E}$ as encoder positions and organize them using a KD-tree based on these positions. Given a set of parameters $\theta$, we traverse the KD-tree to gather the $K$ nearest encoders around $\theta$ and interpolate their weights using inverse distance weighting (IDW), also known as Shepard's algorithm \cite{shepard1968two}: 
\begin{equation}
E(\theta) = 
\begin{cases}
\sum_{j=1}^N \frac{w_j E_j}{\sum_{i=j}^N~w_j} & \text{if } d(\theta, \theta_i) \neq 0,~\forall~j \\
E_j & \text{if } d(\theta, \theta_j) = 0,
\end{cases}
\end{equation}
where $w_j = d(\theta, \theta_j)^{-p}$ with $p=1$. 
If the scene parameter space $\Theta$ is one-dimensional, a fast-path is provided by replacing the KD-tree with a linear array and performing linear interpolation instead of IDW.

Importantly, encoder positions $\mathcal{E}$ can be distinct from the training set $\mathcal{C}$. We combine Bridson's fast Poisson Disk sampling algorithm \cite{bridson2007fast} and Gaussian kernel sampling to generate $\mathcal{E}$. Poisson Disk Sampling ensures even distribution of encoders in the parameter space, maintaining a minimum distance between encoders while preventing regular grid-like patterns. Gaussian kernel sampling enables integration of application-specific knowledge into the network. In \Cref{sec:ablation-params}, we perform an ablation study to analyze the impact of $\mathcal{E}$ on the network's generalization capabilities.

\subsection{Organize Weight Space via Knowledge Distillation}

% Knowledge distillation is the process of transferring knowledge from a large model to a smaller one. 
% It can be viewed as a model compression method that allows a relatively simple model to perform tasks almost as accurately as a very complex model. 

% Knowledge distillation is the process of transferring knowledge from one model to the other. It was first successfully demonstrated by Bucilua~\etal~\cite{bucilua2006model}. This method was later systematically formalized by Hinton~\etal~\cite{hinton2015distilling}. 

% Knowledge distillation is a powerful technique for transferring knowledge from one model to another. Initially demonstrated by Bucilua~\etal~\cite{bucilua2006model}, the approach was later formalized by Hinton~\etal~\cite{hinton2015distilling}. In this work, we employ knowledge distillation to improve HyperINR's performance, using a fully-connected conditional INR model with strong generalizability.

% The process works as the following: First, a fully-connected conditional INR with good generalizability is used as the teacher model $\Omega$ and pre-trained on the training set. Then, a distillation set $\mathcal{D}$ is created by  sampling a set of parameters  $\{ \theta_k \}$ and computing the corresponding data $\mathcal{D} = \{ \Omega(\theta_k), \theta_k \}$ using the trained teacher model. Finally, we complete the knowledge distillation process by optimizing HyperINR using $\mathcal{D}$. A visual illustration of this process can be found in \Cref{fig:eval}.

Knowledge distillation is a powerful method for knowledge transfer between models, with early demonstrations by Bucilu{\u{a}}~\etal~\cite{bucilua2006model} and subsequent formalizations by Hinton~\etal~\cite{hinton2015distilling}. In this study, we leverage knowledge distillation to enhance the performance of HyperINR by employing a fully-connected conditional INR with strong generalizability as the teacher model.
The process begins with training the teacher model, denoted as $\Omega$, on the training set. Then, a distillation set $\mathcal{D}$ is created by strategically sampling a set of scene parameters $\{ \theta_k \}$, and computing the corresponding data $\{ \Omega(\theta_k) \}$ using the trained teacher model. Finally, the HyperINR is optimized using $\mathcal{D}$. A visual illustration of this process can be found in \Cref{fig:eval}.

The distillation set $\mathcal{D}$ can be pre-computed or generated on-demand. Pre-computing $\mathcal{D}$ avoids the training process being bottlenecked by the inference bandwidth of the teacher model, while generating $\mathcal{D}$ on-demand can greatly reduce memory usage. In this work, we pre-compute $\mathcal{D}$.
Furthermore, the selection of a high-quality teacher model and the construction of the distillation set $\mathcal{D}$ can be critical for producing a good HyperINR. In our work, we choose CoordNet as the teacher model due to its remarkable data generation capability.

% \cad{Furthermore, the selection of a high-quality teacher model and the distillation set can be critical for producing a good HyperINR.}

\subsection{Multiresolution Hash Encoding}

\begin{table}[tb]
  \caption{Datasets Used in All Tasks}
  \vspace{-1.0em}
  \label{tab:datasets}
  \scriptsize%
  \centering%
  \begin{tabu}{ccccc}\toprule
    Data & Count & Dimensions & Input/Output & Task \\
    \midrule
    Vortices & 100 & $128,128,128$ & $(t,x,y,z) \mapsto v $ & TSR\\
    Pressure & 105 & $128,128,128$ & $(t,x,y,z) \mapsto v $ & TSR\\
    Temp     & 100 & $864,240,640$ & $(t,x,y,z) \mapsto v $ & TSR\\
    MPAS     & 200 & $256,256$ & $(\theta, \phi, x, y) \mapsto (r,g,b) $ & NVS\\
    MechHand & 150 & $256,256,256$ & $(\theta, \phi, x, y, z) \mapsto v $ & DGS\\
    \bottomrule
  \end{tabu}%
\end{table}

\begin{figure}[tb]
  \centering
  \vspace{-0.5em}
  \includegraphics[width=0.9\linewidth]{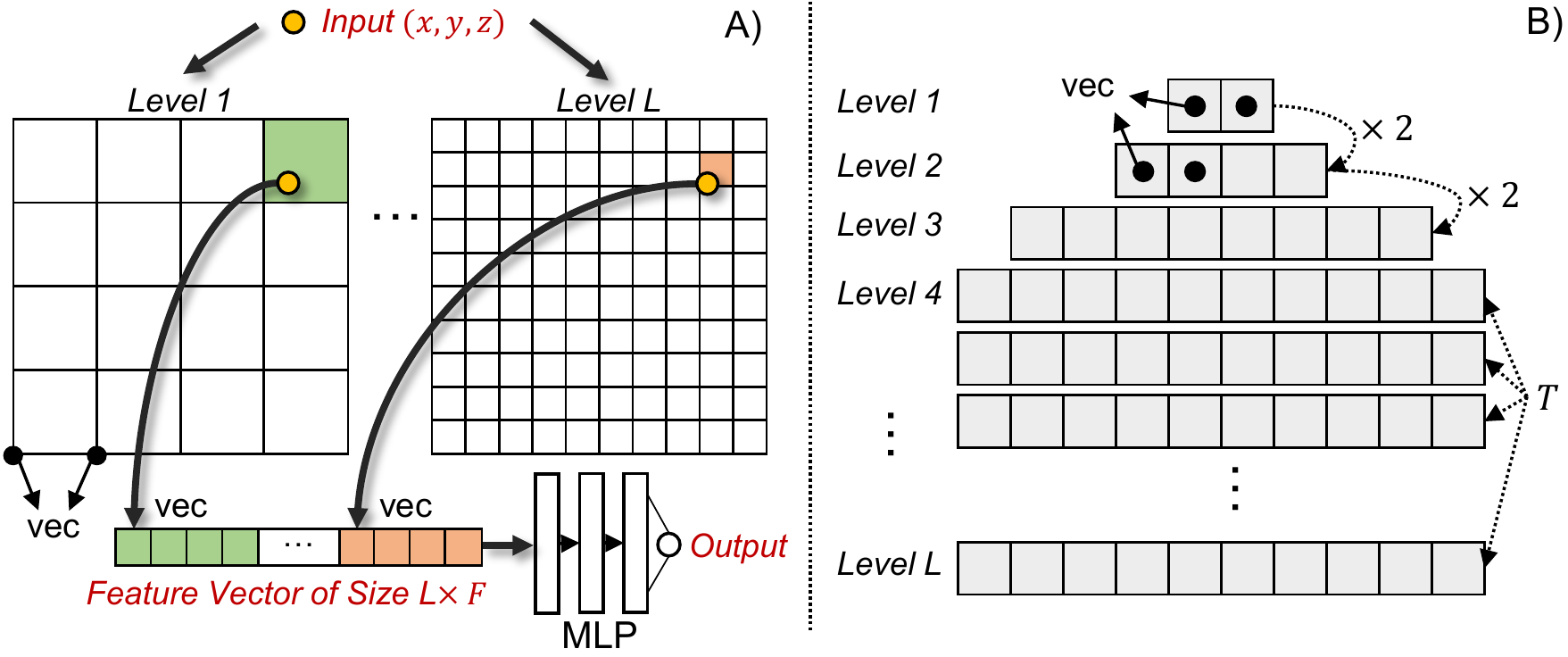}
  \vspace{-1em}
  \caption{Multiresolution Hash Encoding: A) The encoding process and the virtual grids represented by the hash encoding. B) The actual data arrangement within the hash encoding.}
  \label{fig:hash-encoding}
  \vspace{-1em}
\end{figure}

Multiresolution hash encoding~\cite{muller2022instant} is a key technique that enables interactive visualization for our approach. This method models the encoding function $E$
using $L$ levels of independent hash tables with each containing up to $T$ feature vectors of length $F$.
Each level conceptually corresponds to a virtual grid with feature vectors stored at its vertices.
\Cref{fig:hash-encoding}A illustrates the steps performed in the encoding process.

% One of the most important techniques that enables interactive visualization for our technique is multiresolution hash encoding~\cite{muller2022instant}.
% In the following paragraph, we give a detailed explanation of  the inner workings of this technique.
 
% Multiresolution hash encoding models the aforementioned encoding function $E$ that maps an input coordinate $\vec{x}$ to a high dimensional vector.
% % 
% This method arranges additional training parameters into $L$ levels, with each level containing up to $T$ feature vectors of length $F$. \Cref{fig:hash-encoding}A illustrates the steps performed in multiresolution hash encoding. 
% Each level is independent and conceptually corresponding to a virtual grid with feature vectors stored at its vertices.

% The grid resolution $R_l$ starts at a base value $R_{1}$ and increases progressively ($\times 2$ in this paper) as the level $l$ increases.
% Input coordinates are expected to be normalized to $[0, 1]^d$, and then scaled to the grid's resolution.
% $\vec{x_l} = \vec{x} \cdot R_l + 0.5$.
% The offset of 0.5 causes different scales to be staggered with respect to each other, thus preventing spurious alignment of fractional coordinates upon integer scales.
% The output encoding at this level is calculated by interpolating grid vertices based on $x_l$. In this paper, linear interpolation is used.

The grid resolution $R_l$ starts at a base value $R_{1}$ and increases progressively ($\times 2$ in this paper) as the level $l$ increases. Input coordinates are expected to be normalized to $[0, 1]^d$, and then scaled to the grid's resolution: $\vec{x_l} = \vec{x} \cdot R_l + 0.5$.
The offset of 0.5 causes different scales to be staggered with respect to each other, thus preventing spurious alignment of fractional coordinates upon integer scales. The output encoding at this level is calculated by interpolating grid vertices based on $x_l$. In this paper, we use linear interpolation.

\Cref{fig:hash-encoding}B highlights how trainable parameters are stored in the data structure. The grid resolution $R_l$ starts from a relatively small base value, and the number of  vertices at this level might be smaller than $T$. In this case, encoding parameters are directly organized as a linear. When the number of  vertices becomes greater than $T$, a spatial hash function is used to condense encoding parameters.
The hash function used is given by:
\begin{equation}
\text{hash}(\vec{x}) = \left(  \oplus_{i=1}^d x_i \pi_i \right),
\end{equation}
where $\oplus$ is a bitwise XOR operation and $\pi_i$ are unique, large prime numbers. Based on M{\"u}ller~\etal~\cite{muller2022instant}'s recommendation, we use $\pi_1 = 1$, $\pi_2 = 2~654~435~761$, and $\pi_3 = 805~459~861$. Since multiresolution hash encoding is designed to encode spatial coordinates, only 3 prime numbers are used.

% \Cref{fig:hash-encoding}B highlights how trainable parameters are actually stored in the data structure.
% The grid resolution $R_l$ starts from a relatively small base value, and the number of grid vertices at this level might be smaller than $T$. In this case, encoding parameters are directly organized as a 2D or 3D grid.
% When the number of grid vertices becomes greater than $T$, a spatial hash function is used:
% \begin{equation}
% \text{hash}(\vec{x}) = \left(  \oplus_{i=1}^d x_i \pi_i \right),
% \end{equation}
% where  $\oplus$ is a bitwise XOR operation and $\pi_i$ are unique, large prime numbers.
% Based on M{\"u}ller~\etal~\cite{muller2022instant}'s recommendation, we use $\pi_1 = 1$, $\pi_2 = 2~654~435~761$, and $\pi_3 = 805~459~861$.
% Because multiresolution hash encoding is designed to encode spatial coordinates, only 3 prime numbers are used.

\begin{figure}[tb]
  \centering
  % \begin{subfigure}[b]{0.55\columnwidth}
  % 	\centering
  % 	\includegraphics[width=\textwidth]{figs/params_images.pdf}
  % 	\caption{The letter A.}
  % 	\label{fig:ex_subfigs_a}
  % \end{subfigure}%
  % \hfill%
  % \begin{subfigure}[b]{0.45\columnwidth}
  % 	\centering
  % 	\includegraphics[width=\textwidth]{figs/params_shadowmap.pdf}
  % 	\caption{The letter B.}
  % 	\label{fig:ex_subfigs_b}
  % \end{subfigure}%
  \includegraphics[width=\columnwidth]{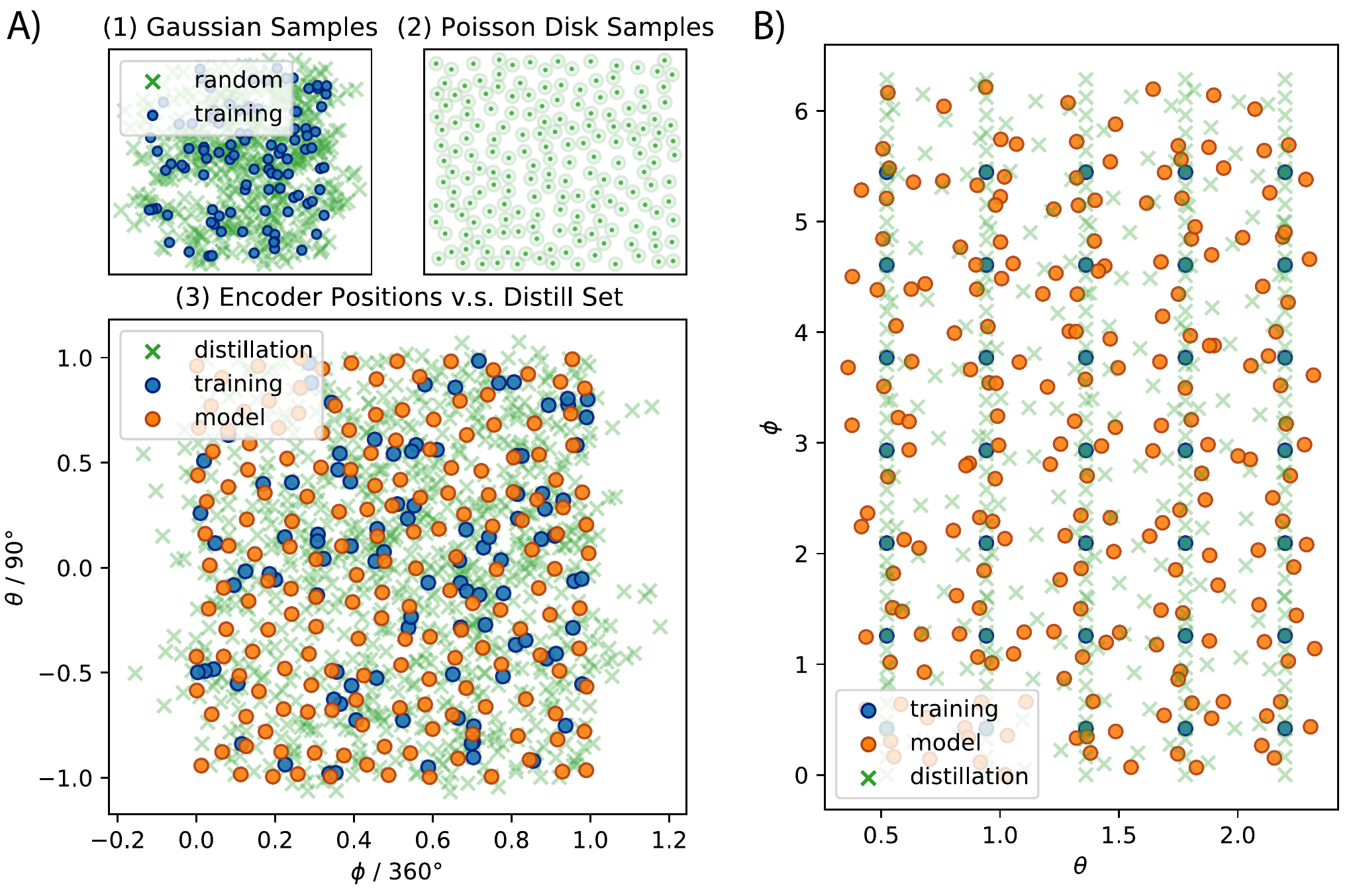}
  \vspace{-2em}
  \caption{A) The visualization of encoder positions $\mathcal{E}$ and the distillation set $\mathcal{D}$ for the novel view synthesis task. We generated encoder positions using Poisson disk sampling (2). We generated distillation set also using Poisson disk sampling, but combined with Gaussian kernel sampling (1). We centered Gaussian kernels on training set. B) The same visualization for the dynamic global shadow task. In this task, we also included some distillation parameters sampled uniformly in the parameter space.}
  \label{fig:params}
  \vspace{-1em}
\end{figure}

% \cad{I am going to attempt to vary this section's sentence structure here. There are too many 'we ...' sentences in-a-row.} :)

\section{Applications}

% In this paper, we apply HyperINR to three generative tasks commonly seen in scientific visualization: novel view synthesis (NVS), temporal super resolution (TSR), and dynamic global shadows for volume rendering (DGS). In this subsection, we describe the setup as well as datasets used for each task. The dataset usages are also summarized in \Cref{tab:datasets}.

In this paper, we apply HyperINR to three common generative tasks in scientific visualization, namely novel view synthesis (NVS), temporal super resolution (TSR), and dynamic global shadows for volume rendering (DGS). Detailed descriptions of setups and datasets used for each task are provided below. In addition, the usages of the datasets are summarized in \Cref{tab:datasets}.

% \cad{Detailed descriptions of setups and datasets used for each task are provided below. In addition, the usages of the datasets are summarized in Table 1.}

% \qwu{Consider describing applications and datasets here?}

\paragraph{\textbf{Novel View Synthesis}}
The objective of NVS is to generate meaningful and visually coherent images of a scene from previously unobserved viewpoints or perspectives, utilizing a collection of pre-existing images. NVS holds significant potential in the realm of scientific visualization, enabling the creation of explorable images~\cite{tikhonova2010explorable}. To perform this task, an INR should accept two spatial inputs $(x, y)$ and a viewing direction, subsequently producing an RGB color $(r, g, b)$. In this study, we parameterize the viewing direction using a spherical coordinate system, characterized by a polar angle $\theta$ and an azimuthal angle $\phi$.
Our experiment employs a dataset of 200 isosurface visualizations, generated by He~\etal~\cite{he2019insitunet} utilizing the MPAS-Ocean simulation from the Los Alamos National Laboratory, referred to as the \textbf{MPAS} dataset. 
100 of the visualizations were allocated for training, and the remaining were for testing purposes. Furthermore, the quality of the synthesized images was assessed through the Peak Signal-to-Noise Ratio (PSNR) metric. The model's inference performance was evaluated in terms of the data bandwidth.

\paragraph{\textbf{Temporal Super Resolution}}
The goal of TSR is to train a neural network on a sequence of sparsely sampled time-varying volume data and enable the generation of the same sequence at a higher temporal resolution. A more complete review of related TSR techniques can be found in~\Cref{sec:related-generative-model}.
% 
% \cad{Confusing. The focus of our TSR task is on scalar field volume data, where...}
We focus on scalar field volume data, where an INR receives a 4D input $(x, y, z, t)$ and outputs a scalar value $v$. Although all of our experiments assumes $v$ to be a scalar value, our method can be extended to multivariate volume data without any alterations to the design. We assess our approach using three datasets: 1) a time-varying simulation of \textbf{vortices} provided by Deborah Silver at Rutgers University, consisting of 100 timesteps with 20 equally spaced steps used for training; 2) the \textbf{pressure} field of a Taylor-Green Vortex simulation generated by the NekRS framework, containing 105 timesteps with 21 selected for training; and 3) the \textbf{temperature} field from a 1atm flame simulation produced by S3D~\cite{s3d}, which includes 90 timesteps with 10 employed for training.
To evaluate the quality of the generated data, we employ metrics such as PSNR and the Structural Similarity Index Measure (SSIM). For assessing the inference performance, we measure the average inference bandwidth as well as the interactive volume visualization framerates.

\paragraph{\textbf{Dynamic Global Shadows}}
Direct volume rendering with global shadows is a non-physically based shading technique widely employed in scientific volume rendering. It enhances realism and helps distinguish features in the data. However, the technique also imposes a significant runtime cost over simple ray casting, as it generates at least one secondary ray towards the light source at each sample point to estimate shadow coefficients. This results in an $O(n^2)$ computation for each primary ray. 
An alternative approach involves precomputing all secondary rays at voxel centers, generating a volume data containing shadow coefficients.
This ``shadow volume'' is subsequently utilized to estimate shadow coefficients at sample positions. Although this method reduces the computational complexity to $O(n)$ per ray, it substantially increases the memory footprint for rendering. Furthermore, the shadow volume must be regenerated whenever the transfer function or the light changes, presenting challenges for interactive exploration.

In this work, as a preliminary study, we examine the potential of achieving dynamic global shadows using HyperINR. Specifically, we propose substituting shadow volumes with regular INRs and estimating shadow coefficients through network inferences. We term these INRs as shadow INRs. This optimization significantly reduces memory footprints. Then, we optimize a HyperINR to generate such INRs. The resulting HyperINR can achieve dynamism for global shadows as the generation process can be done in real-time.
We validate this method by generating a set of 150 shadow volumes sampled with varying light positions. For this preliminary study, we fix the transfer function and incorporate only one light source. Then, we utilize 35 evenly distributed shadow volumes to optimize our network. To assess the shadow generation quality, we computed the PSNR and SSIM against the ground truth data. 
To evaluate the inference performance, we measured the rendering framerate and INR generation latency.

\section{Implementation}\label{sec:network-impl}

Our network is implemented in PyTorch, with GPU-accelerated training using the Tiny-CUDA-NN machine learning framework~\cite{tiny-cuda-nn}. We leverage multiresolution hash encoding for training the INR.

\paragraph{\textbf{Architecture}}
% For all tasks, a base hash encoding unit consists of $L=8$ encoding levels. Each level contains up to $T=2^{15}$ feature vectors of size $F=4$. 
% These hyperparameters are chosen based on results shown in \Cref{sec:ablation-params}.
% We set the base grid resolution to be $R_1=8$ for image-based applications and $R_1=4$ for volume-based applications. 
HyperINR's base hash encoder consists of $L=8$ encoding levels, with each level containing up to $T=2^{15}$ feature vectors of size $F=4$. We selected these hyperparameters based on the results presented in \Cref{sec:ablation-params}. The base grid resolution is set to $R_1=8$ for the NVS task and $R_1=4$ for TSR and DGS tasks.
% 
% Next, for the NVS task, we randomly generated 177 encoder positions using Poisson disk sampling in a $[0,1]^2$ with radius 0.06.
% For the TSR task, we distribute 24 encoding units evenly across the temporal domain.
% For the DGS task, we created 206 encoder positions using both Poisson disk sampling and  Gaussian distribution sampling.
% We study the effect of hyperparameter in \Cref{sec:ablation-params}.
% 
The MLP unit adopts the configuration proposed by Wu~\etal~\cite{wu2022instant}, which uses four hidden layers and a width of 64 neurons. This configuration is suitable for the volume rendering algorithm.

As for encoder positions $\mathcal{E}$, 
% For the NVS task, we randomly generated 177 encoder positions using Poisson disk sampling within a $[0,1]^2$ space with a radius of 0.06. 
177 encoder positions were randomly generated using Poisson disk sampling within a $[0,1]^2$ space for the NVS task,
% 
% we distribute 24 encoding units evenly across the temporal domain. 
24 encoding units were evenly distributed across the temporal domain for the TSR task,
% 
% we created 206 encoder positions using both Poisson disk sampling and Gaussian distribution sampling. 
and 206 encoder positions were created using both Poisson disk sampling and Gaussian distribution sampling for the DGS task.
The impact of these hyperparameters is further explored in \Cref{sec:ablation-params}.

\paragraph{\textbf{Distillation}}
We select CoordNet as the teacher model for knowledge distillation. However, since a reference implementation of CoordNet is not publicly available, we implement it based on SIREN~\cite{sitzmann2019siren} and NeurComp~\cite{lu2021compressive}. Our implementation closely matches the architecture described in the CoordNet paper~\cite{han2022coordnet}: we use 3 resblocks as the encoder to process inputs, 10 hidden resblocks of size 256, and 1 resblock as the decoder to produce final outputs.
It is worth noting that CoordNet expects input and output values to be within the range of $[-1,1]$, whereas hash-encoding-based INRs use a range of $[0,1]$. Therefore, in our implementation, we use $[0,1]$ as the value range and only convert it to the $[-1,1]$ range when interacting with CoordNet.

\paragraph{\textbf{Training}} 
% We develop an end-to-end training framework to optimize our hypernetwork instead of working on pre-trained network weights. To achieve this, we implement a completely stateless INR evaluator (as shown in \Cref{fig:eval}) that can take a predicted network weights $W$ and a pre-calculated coordinate matrix $C$ to predict  output data $V_{\text{pred}}$. Then the loss between $V_{\text{pred}}$ and the corresponding ground truth data $V_{\text{true}}$ is computed. The INR evaluator can calculate gradients with respect to $W$ and back-propogate them to the hypernetwork.
% 
We develop an end-to-end training framework to optimize our hypernetwork instead of relying on pre-trained network weights. To achieve this, we implement a stateless INR evaluator (as shown in \Cref{fig:eval}) that takes a coordinate matrix $C$ and a network weight matrix $W$ predicted by the hypernetwork as inputs, and generates output data $V_{\text{pred}}$. A loss is thus calculated  with respect to the ground truth $V_{\text{true}}$. Then the evaluator can compute gradients with respect to $W$ and backpropagate them to the hypernetwork.
For NVS, we use the $\mathcal{L}^2$ loss between pixel colors. For TSR and DGS, we use the $\mathcal{L}^1$ loss following the recommendations of Han~\etal~\cite{han2022coordnet}.
% For explorable images, the $\mathcal{L}^2$ loss between pixel colors is used. For temporal super resolution and dynamic volume shadows, the $\mathcal{L}^1$ loss is used following recommendations by Han~\etal~\cite{han2022coordnet}.

% To optimize HyperINR, we use the Adam optimizer and set $\beta_1 = 0.9$, $\beta_2 = 0.999$, $\epsilon = 10^{-10}$. We observe good convergence speed with a learning rate of $10^{-3}$. For the teacher model, CoordNet, we also use the Adam optimizer but set $\beta_1 = 0.9$, $\beta_2 = 0.999$, $\epsilon = 10^{-8}$ and a weight decay $10^{-6}$. We find that CoordNet can be very sensitive to learning rate, and a learning rate of $10^{-5}$ can enable stable convergence. For the NVS task, we optimized CoordNet for 300 epochs, and then created a distillation set $\mathcal{D}$ with 981 samples using a combination of Poisson disk sampling and Gaussian kernel sampling, similar to the generation of encoder positions. $\mathcal{D}$ is visualized in \Cref{fig:params}A. For TSR tasks, we trained CoordNet for 30k epochsto ensure that the teacher model is well-optimized. Then we created distill samples by uniformly sampling the time axis. We summarize different datasets' $\mathcal{D}$ sizes in \Cref{tab:results}. For the DGS task, we also trained a CoordNet for 30k epochs, then generated a $\mathcal{D}$ containing 400 samples, visualized in \Cref{fig:params}B.

% \cad{... and epsilon = ..., A good convergence speed was observed with a learning rate of 10e-3. The teacher model, CoordNet, was also trained using the Adam optimizer with...}

To optimize HyperINR, we use the Adam optimizer with $\beta_1 = 0.9$, $\beta_2 = 0.999$, and $\epsilon = 10^{-10}$. A good convergence speed was observed with a learning rate of $10^{-3}$. The teacher model, CoordNet, was also trained using the Adam optimizer with $\beta_1 = 0.9$, $\beta_2 = 0.999$, $\epsilon = 10^{-8}$, and a weight decay of $10^{-6}$. We find that CoordNet can be sensitive to the learning rate and a learning rate of $10^{-5}$ enables stable convergence. For NVS, we optimize CoordNet for 300 epochs and then create a distillation set $\mathcal{D}$ using a combination of Poisson disk sampling and Gaussian kernel sampling, similar to the generation of encoder positions. The $\mathcal{D}$ is visualized in \Cref{fig:params}A. For TSR tasks, we train CoordNet for 30k epochs to ensure good performance for the teacher model. Then, we create distillation sets by uniformly sampling the time axis.  For the DGS task, we also train CoordNet for 30k epochs but generate a $\mathcal{D}$ containing 400 samples, which is visualized in \Cref{fig:params}B. 
% We summarize the sizes of $\mathcal{D}$ in \Cref{tab:results}.
\Cref{tab:results} contains the summary of the sizes of the distillation set $\mathcal{D}$.

% \cad{Table 2 contains the summary of the sizes of the distillation set D.}

We utilize automatic mixed precision~\cite{micikevicius2017mixed} to accelerate training and reduce the memory footprint. To avoid data underflow in the backpropagation process, gradient scaling was also employed. Some of the experiments were conducted on the Polaris supercomputer at the Argonne Leadership Computing Facility.

% \cad{...To avoid data underflow in the backprop process, gradient scaling is also used.}

% To accelerate training and reduce memory footprint, we 
% also applied automatic mixed precision training~\cite{micikevicius2017mixed} to use half precision floating point numbers if possible. To avoid data underflow in the back propagation process, we also employed gradient scaling. 
% We performed some of the experiments on the Polaris supercomputer from the Argonne Leadership Computing Facility.

\paragraph{\textbf{Initialization}} 
Properly initializing network weights is crucial for both HyperINR and CoordNet. For HyperINR, we follow M{\"u}ller~\etal's suggestions and initialize the hash table entries using the uniform distribution $\mathcal{U}(-10^{-4}, 10^{-4})$~\cite{muller2022instant}. This approach provides a small amount of randomness while encouraging initial predictions close to zero, allowing hash encoding units to converge properly.
CoordNet heavily utilizes SIREN layers, so we apply SIREN's initialization scheme and use the uniform distribution $\mathcal{U}(-\frac{\sqrt{6/n}}{30}, \frac{\sqrt{6/n}}{30})$ to initialize CoordNet weights, with $n$ being the number of neurons in the layer~\cite{sitzmann2019siren}. 
% This approach has been shown to be effective in initialization of SIREN layers.

% Properly initializing network weights can be key for both HyperINR and CoordNet. For HyperINR, in order to allow hash encoding units to converge properly, we have to strictly follow M{\"u}ller~\etal's suggestions and initialize initialize the hash table entries using the uniform distribution $\mathcal{U}(-10^{-4}, 10^{-4})$ to provide a small amount of randomness while encouraging initial predictions close to zero. Because CoordNet heavily uses SIREN layers internally, we apply SIREN's initialization scheme~\cite{sitzmann2019siren} and uses the uniform distribution $\mathcal{U}(-\frac{\sqrt{6/n}}{30}, \frac{\sqrt{6/n}}{30})$ with $n$ being the number of neurons in layer.

% \cad{For the implementation part, I think the current spec (hyper-parameters here, architecture in the above sections) should suffice, with fig 2 and fig 3 giving enough context on how it works, but personally, albeit nitpicking, the information presented here isn't explicit enough for me to build the network from the paper. For a presentation style and ease of implementation, I would probably diagram the entire model out with UML or others with additional notes that can reference back to explanations in this paper.}

\begin{figure}[tbp]
  \centering
  \includegraphics[width=\linewidth]{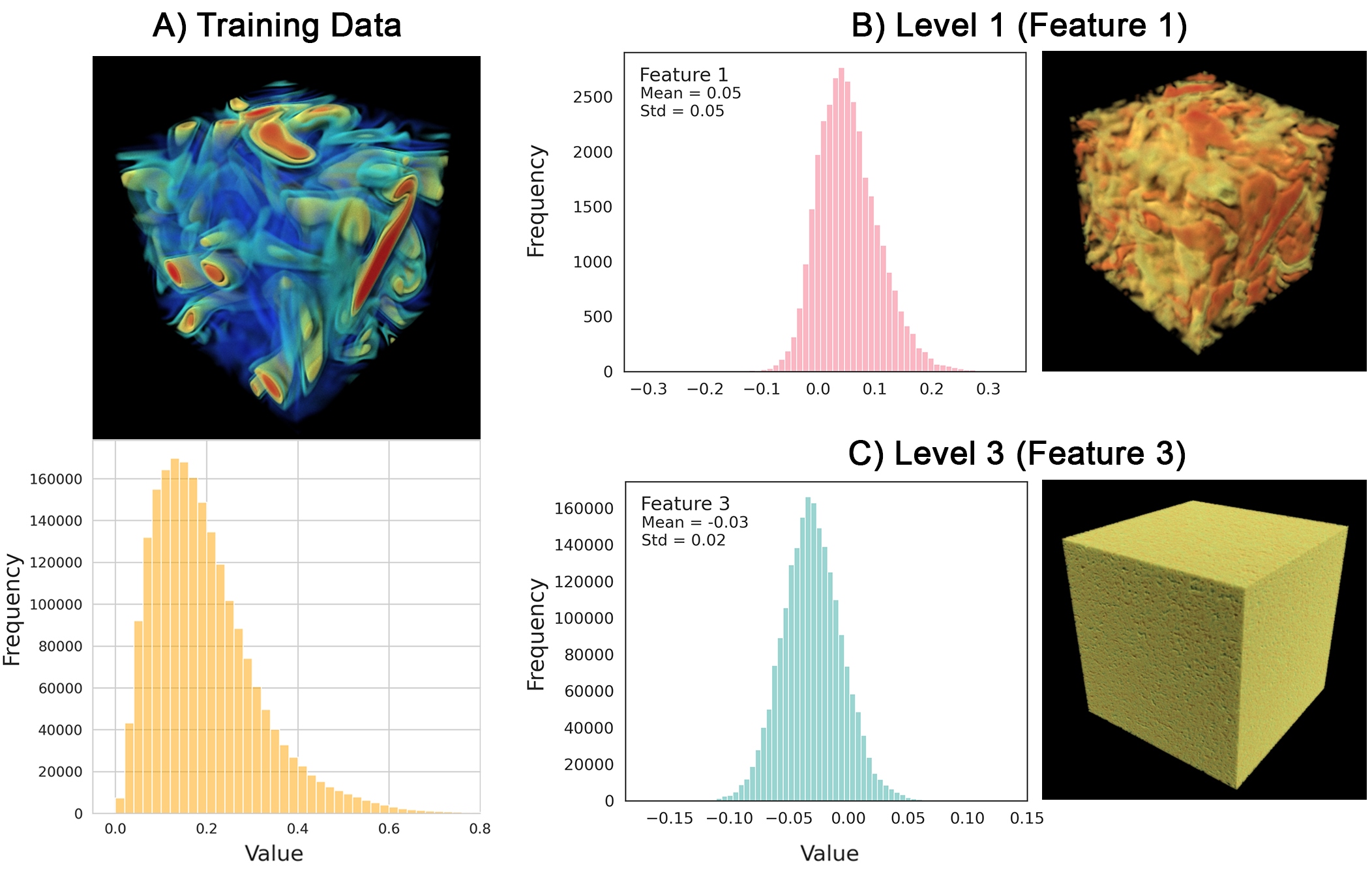}
  \vspace{-2em}
  \caption{A) The volume visualization of the \textbf{vortices} data as well as the histogram data values. B), C) Volume visualizations and histograms of the corresponding hash encoding weights.}
  \label{fig:histogram}
  \vspace{-1em}
\end{figure}

\begin{figure*}[tbp]
  \centering
  \vspace{-0.5em}
  \includegraphics[width=\linewidth]{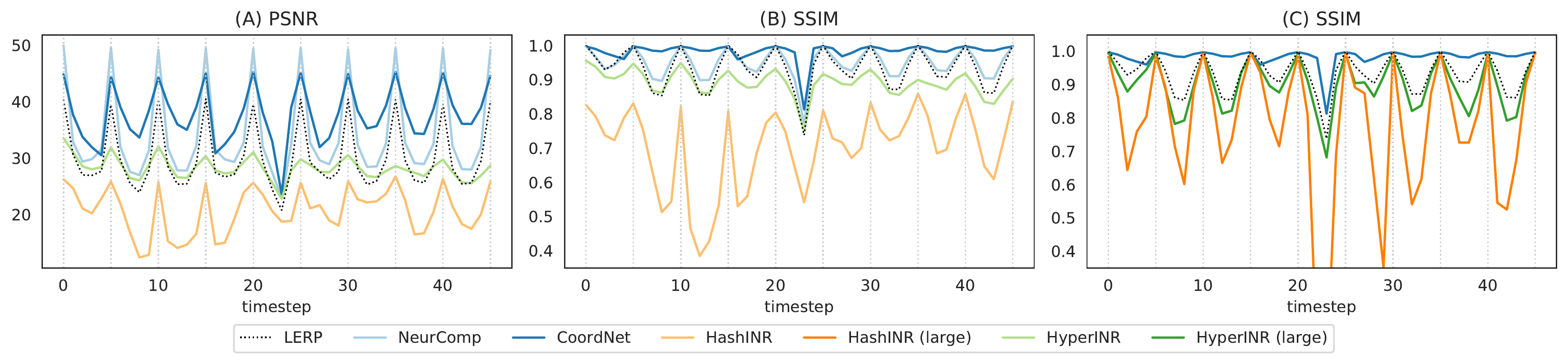}
  \vspace{-2em}
  \caption{Comparisons of different INR architectures on the temporal super-resolution task. A), B) Comparisons of networks with equalized number of trainable parameters. C) To rule out the possibility that HashINR and HyperINR were limited by network capacity, another comparison were performed against a HashINR with a very large hash table size ($T = 2^{21}$) and a HyperINR with sufficient hash encoders ($N = 24$).}
  \label{fig:diffinrs}
  \vspace{-1.2em}
\end{figure*}

\section{Ablation Study}

In this section, we present results and findings that motivate the design of HyperINR, and conduct a hyperparameter study to determine the optimal configuration for our tasks.

\subsection{Understand Hash Encoding Weights}

% To design a hypernetwork that can properly predict multiresolution hash encoding weights, we need to first understand these weights. To do this, we first remapped each encoding level's weights back to the corresponding grid space: $\{~g_{x,y,z} = W_{\text{hash}(x,y,z)}~\}$. Then, we computed histograms and volume renderings of the weight grid, level by level and feature by feature. We show visualization results in \Cref{fig:histogram}.

% To design a hypernetwork that accurately predicts multiresolution hash encoding weights, it is crucial to first comprehend these weights. To achieve this, we mapped each encoding level's weights back to the corresponding grid space: $\{~g_{x,y,z} = W_{\text{hash}(x,y,z)}~\}$. Then, we computed histograms and volume renderings of the weight grids, level by level and feature by feature. The visualization results are presented in \Cref{fig:histogram}.

Designing an accurate hypernetwork for predicting multiresolution hash encoding weights requires a thorough understanding of these weights. To gain insight, we mapped the weights of each encoding level back to the corresponding grid space $\{~g_{x,y,z} = W_{\text{hash}(x,y,z)}~\}$, and visualized them using histograms and volume renderings, as shown in \Cref{fig:histogram}.

\paragraph{\textbf{Parametric Encoding is an Embedding of Local Features}}
Our analysis reveals that when the number of grid vertices is smaller than the hash table size, a strong correlation exists between the encoder parameter values and the actual data values. Furthermore, different features tend to capture different details, resulting in distinct weight distributions. These observations suggest that INRs with parametric encoding  operate differently from those parameterized solely by MLPs. With parametric encoding, local details in the data are simply projected into a high-dimensional weight space, rather than being approximated indirectly through MLPs. These findings agree with our intuition about parametric encoding and explain why employing multiresolution hash encoding can result in a loss of generalizability.

% 
% We found that when the total number of grid vertices is less than the hash table size, there is a strong correlation between hash parameter values and the actual data values. Different features seems to capture different details, thus they might result in different weight distributions.
% 
% 
% It indicates INR with multiresolution hash encoding, or perhaps parametric encoding in general, works fundamentally different from INRs parameterized fully by MLPs.
% 
% With parametric encoding, local details presented in the data are simply projected to a high dimensional space, rather than being indirectly approximated by non-linear functions represented by the MLPs. This validates our intuition about parametric encoding, and explains why the use of such encoding method leads to loss in generalizability.

\paragraph{\textbf{Hash Function Breaks Local Similarities}}
When the total number of grid vertices exceeds the hash table size, a spatial hash function is utilized to condense encoding parameters. Our results indicate that this process breaks the aforementioned correlation between parameter values and data values. These local similarities are broken in the hashed space and cannot be restored. This suggests that when designing a hypernetwork for multiresolution hash encoding, levels processed by the spatial hash function must be treated differently from  those without it. Operations such as CNNs that take advantage of local spatial coherence may not work even after remapping encoding parameters back to the grid space. Our findings motivate us to focus on weight interpolation methods that rely solely on correlations between parameters stored at the same location in the hash table.

% When the total number of grid vertices exceeds the hash table size, a spatial hash function is used to condense the encoding features. We found that this process also breaks the correlation between parameter values and data values. This indicates that to design a hypernetwork that will potentially work with multiresolution hash grid encoding, levels that are processed by the spatial hash function should be handled differently than levels without it. Methods that relies on spatial consistencies perhaps will not work even after remapping encoding parameters back to the grid space.
% Local similarities is broken in the hashed space, and cannot be recovered.
% This motivates us to focus on weight interpolation methods which only utilizes correlations between parameters stored at the same location in the hash table.

% \cad{... values and data values, and local similarities would be lost in the hash space, and cannot be restored...}

\begin{figure}[tbp]
  \centering
  \includegraphics[width=\linewidth]{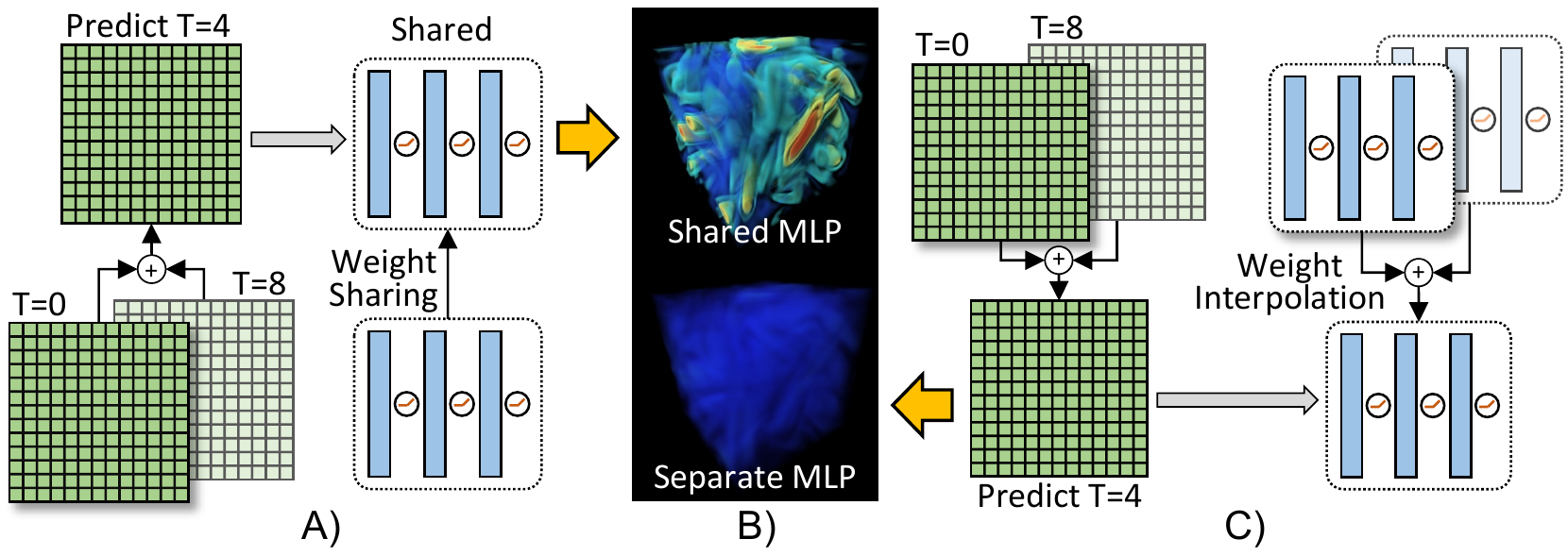}
  \vspace{-2.1em}
  \caption{Comparisons of different weight interpolation methods. A) The method that uses a shared MLP. B) Volume rendering results. C) The na{\"i}ve design that employs different MLPs for different hash encoders.}
  \label{fig:shared-simple}
  \vspace{-1em}
\end{figure}

\subsection{Shared MLP Unit}\label{sec:ablation-shared-mlp}

% The core concept behind HyperINR is to properly organize the weight space of INRs to enable weight interpolation. To this end, we compare two designs for implementing weight interpolation, as visualized in \Cref{fig:shared-simple}.

The fundamental principle behind HyperINR is the effective organization of the INR weight space to facilitate interpolation. Thus, it is crucial to first identify the appropriate interpolation method. In this section, we examine two weight interpolation designs. These two designs are visually illustrated in \Cref{fig:shared-simple}.

% For this scenario, we chose to examine two designs for implementing weight interpolation. These two designs are illustrated in \Cref{fig:shared-simple}.

% \cad{...facilitate weight interpolation. For this scenario, we chose to examine two designs...}

% The simplest design would directly optimize an standalone INR for each training sample, and then interpolate all the network weights for parameters outside the training set. 
% However, we found that such a design is not robust to perturbations (\ie different random seedings), and training over the same dataset may yield vastly different INR weights. 
% 
% These INRs will produce similar inference results, but it is in general meaningless to interpolate among their network weights, because the same network parameter might have totally different meanings in different INR instances. We show results of this design in the lowe part of \Cref{fig:shared-simple}B.

The straightforward design involves optimizing a standalone INR for each training data, and then interpolating all the network weights for parameters outside the training set. However, this design demonstrates a lack of robustness to perturbations (\ie different random seedings), resulting in totally different network weights if repeatedly trained on the same dataset. While these INRs can yield similar inference outcomes, interpolating among their network weights is generally meaningless because network parameters in the same relative location may hold entirely different meanings across INR instances. The lower part of \Cref{fig:shared-simple}B presents a volume rendering result of the INR produced by this weight interpolation design.

% Alternatively, we construct our HyperINR by maintaining a MLP shared among all positional encoders. This MLP acts as a common projector, ensuring that a network parameter has similar meanings across different INR instances, making weight interpolation reasonable. Our results, depicted in the upper part of \Cref{fig:shared-simple}B, confirm the effectiveness of the shared MLP unit in enabling meaningful weight interpolation.

In contrast, we construct our HyperINR by incorporating a shared MLP among all positional encoders. This MLP serves as a common projector, ensuring consistent meanings for network parameters across different INR instances and thus rendering weight interpolation feasible. Our finding, depicted in the upper part of \Cref{fig:shared-simple}B, substantiates the effectiveness of a shared MLP unit  in enabling meaningful weight interpolation for HyperINR.

% The core idea of HyperINR is to properly organize the weight space of INRs so that it become meaningful to perform weight interpolation. In this section, we compare two designs for implementing weight interpolation, visualized in \Cref{fig:shared-simple}.

% \subsection{Comparison with Hypernetworks}
% \label{sec:ablation-hypernet}

\begin{figure*}[tb]
  \centering
   \includegraphics[width=0.98\linewidth]{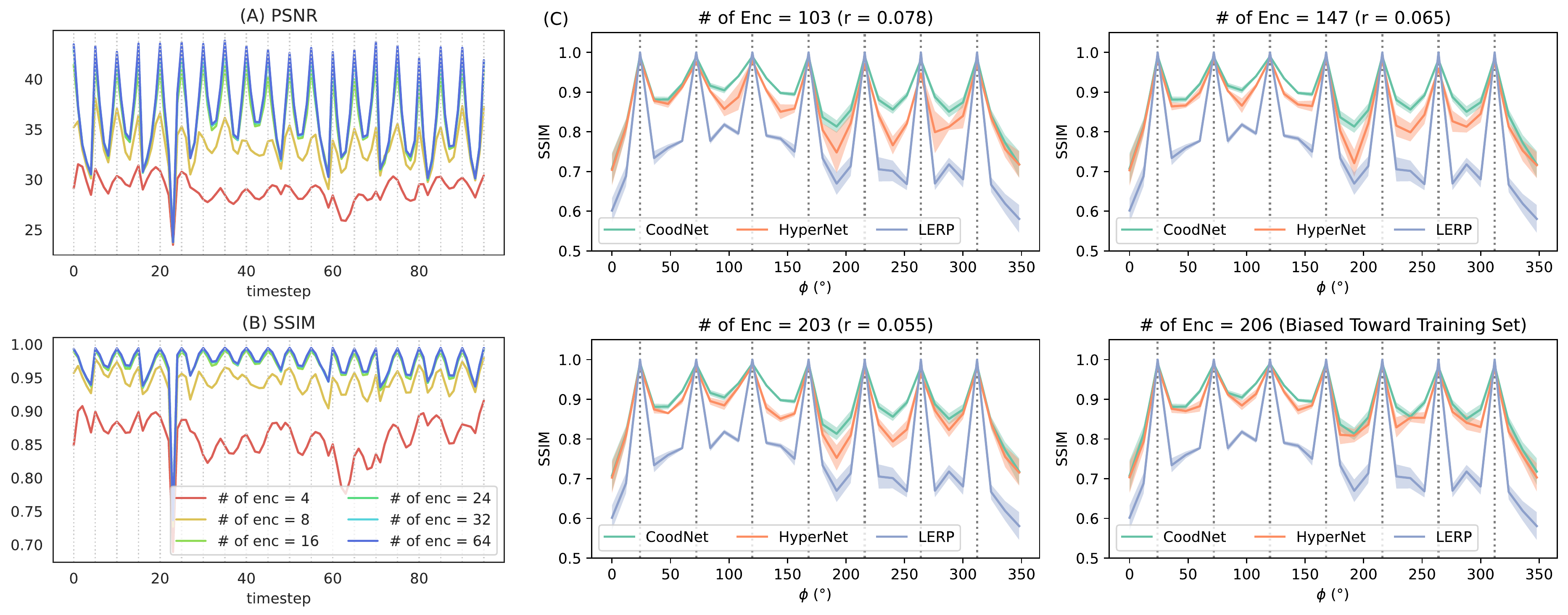}
  % \begin{subfigure}[b]{0.324\linewidth}
  % 	\centering
  %       \includegraphics[width=\textwidth]{figs/view_diffgrids}
  % 	\caption{\qwu{1D case}.}
  % 	\label{fig:diffgrids_1d}
  % \end{subfigure}%
  % \hfill%
  % \hspace{0.1em}
  % \begin{subfigure}[b]{0.668\linewidth}
  % 	\centering
  %       \includegraphics[width=\textwidth]{figs/shadowmap_vis}
  % 	\caption{\qwu{High dimensional case}.}
  % 	\label{fig:diffgrids_2d}
  % \end{subfigure}%
  \vspace{-1.2em}
  \caption{The impact of hash encoder positions $\mathcal{E}$ on the data reconstruction quality was examined in two tasks. The first task, A-B), conducted temporal super-resolution using the \textbf{vortices} dataset. The second task, C), used the \textbf{mechhand} dataset to address dynamic global shadows. In C), each line represents the average SSIM across 5 different $\theta$ angles, with filled areas highlighting the $\pm$ 1 standard deviation regions.}
  \label{fig:diffgrids}
  \vspace{-0.85em}
\end{figure*}

\subsection{Comparison between INRs}\label{sec:ablation-canilla}

% \qwu{equalize parameters, compare neurcomp, coordnet, hash+time large MLP, ours}
% The next question is 

% After identifying a good architecture for HyperINR, we now study whether employing HyperINR is necessary.

With a good interpolation strategy, we now investigate the necessity of employing a hypernetwork to achieve both good generalizability and inference speed. In this section, we present experimental comparison between HyperINR and three distinct INR architectures. We conducted our experiments on the TSR task, utilizing the \textbf{vortices} dataset.

% In this section, we present an experimental comparison between four different INRs to investigate the necessity of building a hypernetwork to achieve both good generalizability and inference speed. We conducted our experiments on the TSR task, using the \textbf{vortices} dataset.

The three neural networks we employed in our experiment are CoordNet~\cite{han2022coordnet}, NeurComp~\cite{lu2021compressive}, and a hash encoding based INR extended to 4 dimensions (HashINR). We included CoordNet and NeurComp in our experiment as CoordNet has recently demonstrated strong performance on TSR tasks, and there is no report of NeurComp on TSR tasks to the best of our knowledge. HashINR processes spatial coordinates using hash encoding and time using OneBlob encoding~\cite{muller2019neural}. Such a HashINR has been used by several computer graphics works~\cite{muller2021real} and demonstrated good learning capabilities while offering very good inference performance.

To ensure fair comparisons, we used the standard CoordNet configuration described in \Cref{sec:network-impl} and equalized the number of trainable parameters used in all the other networks (to around 2.1$\sim$2.2M). Specifically, for NeurComp, we used 10 resblocks of size 327; for HyperINR, we constructed 6 small hash encoders ($T=2^{14}$); and for HashINR, we used medium-sized hash tables ($T=2^{17}$). We optimized all the networks sufficiently and report their performances in \Cref{fig:diffinrs}A-B. Notably, we observed that HashINR did not perform well. To rule out the possibility that the HashINR was limited by its network capacity, we performed another comparison against a HashINR with very large hash tables ($T=2^{21}$) and a HyperINR with sufficient hash encoders ($N=24$), with results highlighted in \Cref{fig:diffinrs}C. 

Our experimental results lead us to three conclusions. Firstly, under equal conditions, CoordNet performed significantly better than NeurComp in terms of generalizability. This is likely due to the differences in network design, as their network capacities were equalized. Secondly, compared with two pure MLP-based INRs, HashINR was unable to achieve good performance using the same amount of trainable parameters. Even after increasing the hash table size, HashINR still struggled to perform well on unseen parameters. Finally, we found that HyperINR outperformed HashINR in terms of generalizability by splitting a larger hash table into multiple smaller ones. However, there was still a performance gap compared to CoordNet. To bridge this gap, we utilize knowledge distillation.

\begin{figure}[tb]
  \centering
  \includegraphics[width=\linewidth]{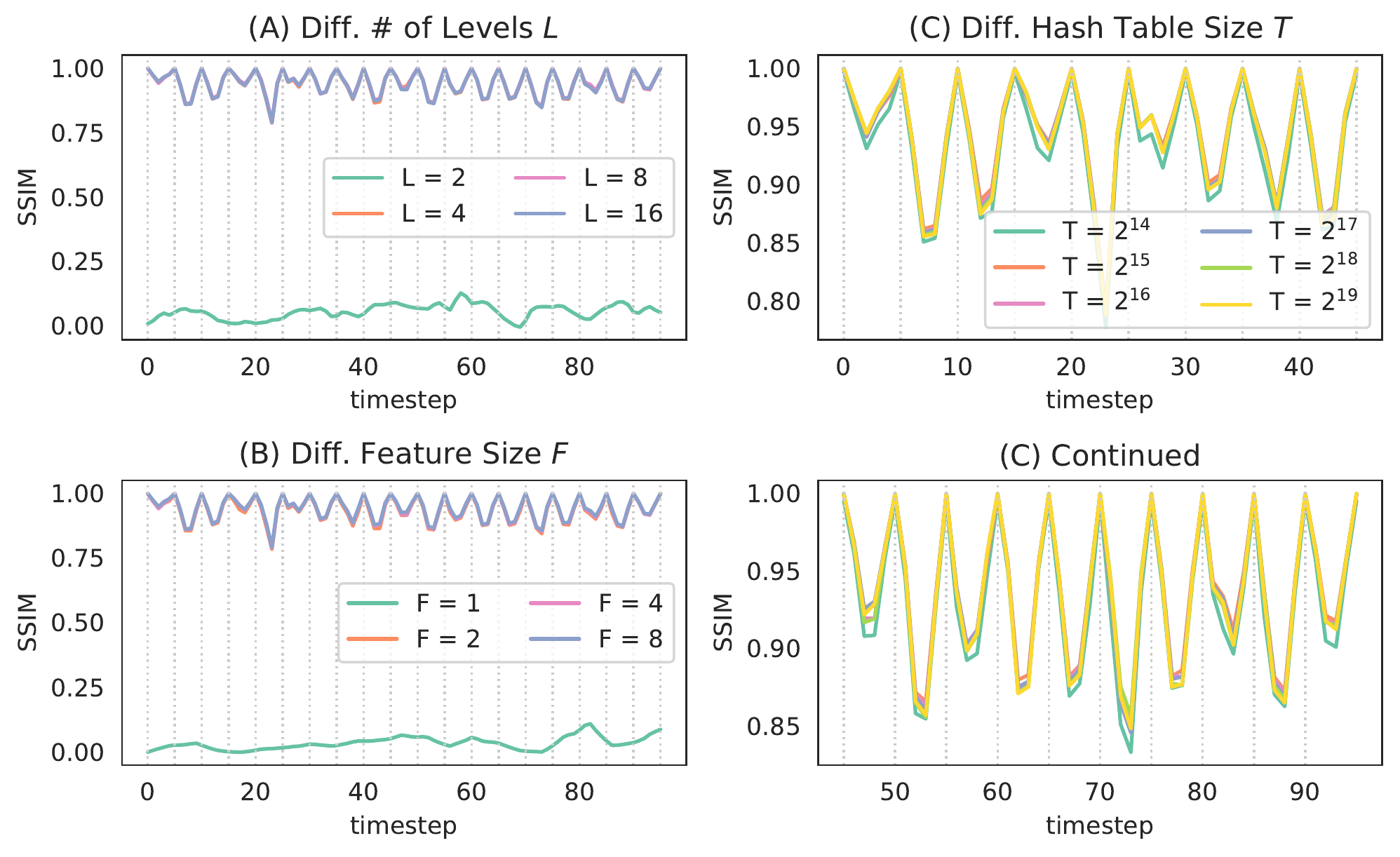}
  \vspace{-2.2em}
  \caption{Hyperparameter study of hash encoder parameters.}
  \label{fig:diffparams}
  \vspace{-1em}
\end{figure}

\subsection{Hyperparameter Study}\label{sec:ablation-params}
% \qwu{Compare different encoding resolution}

% \begin{figure}[tbp]
%   \centering
%   \begin{subfigure}[b]{0.45\columnwidth}
%   	\centering
%   	\includegraphics[width=\textwidth]{figs/params_shadowmap.pdf}
%   	\caption{The letter B.}
%   	\label{fig:ex_subfigs_b}
%   \end{subfigure}%
%   \hfill%
%   \begin{subfigure}[b]{0.55\columnwidth}
%   	\centering
%   	\includegraphics[width=\textwidth]{figs/params_images.pdf}
%   	\caption{The letter A.}
%   	\label{fig:ex_subfigs_a}
%   \end{subfigure}%
%   \caption{TODO.}
% \end{figure}

% In this section, we discuss experiments we performed to find the best hyperparameters for HyperINR. We first study parameters used to describe an individual hash encoder. Then we look into how the choice of hash encoder locations in the parameter space can affect performances.

With a strong network architecture, we now determine the optimal hyperparameters for HyperINR. Specifically, we examined parameters related to individual hash encoders, as well as the impact of encoder positions in the parameter space on performance.

\paragraph{\textbf{Hash Encoder Parameters}}\label{sec:hash-encoder-param}
To evaluate parameters related to individual hash encoders, we constructed various HyperINRs and tasked them with performing TSR on the \textbf{vortices} dataset. We maintained a fixed number of 24 encoders, distributed uniformly throughout the temporal domain, and assessed reconstruction quality using SSIM. Results are reported in \Cref{fig:diffparams}.

First, we set $T=2^{15}$, $F=4$, and varied the number of encoding levels $L$ from 1 to 16. Our findings suggest that $L>2$ is sufficient for this particular TSR problem. Increasing $L$ does not yield significant improvements in generalizability.
Next, we held $L$ constant at 8 and varied the number of features per encoding level $F$ from 1 to 8. Results showed that good performance can be achieved with $F>1$, but increasing $F$ yields diminishing returns.
Finally, with $F=4$ held constant, we adjusted the hash table size $T$ from $2^{14}$ to $2^{19}$. In general, we observed that a larger hash table size leads to better reconstruction quality, although the differences were minimal for our particular problem. 

We conclude that selecting appropriate hyperparameters depends on the data complexity, and for more complicated data, larger hyperparameter values are likely necessary. We repeated these experiments using other datasets and determined that $T=2^{15}$, $F=4$, and $L=8$ generally produce good performance across all cases considered in this study.

% We construct different HyperINRs to perform the same TSR task on \textbf{Vortices} data. For all runs, we fixed the number of encoders to be 24 and uniformly distribute them along the temporal domain. We measure reconstruction quality using the structural similarity metrics (SSIM). Results are reported in \Cref{fig:diffparams}. 
% First, we set $T=2^{15}$, $F=4$ and vary the number of encoding levels $L$ from 1 to 16. We found that for this TSR problem, $L>2$ was sufficient. Increasing $L$ cannot significantly improve generalizability. 
% Next, we fixed $L=8$, and vary the number of features per encoding level $F$ from 1 to 8. We observed very similar results where good performance can be achieve with $F>1$. Keep increasing $F$ produced diminishing returns. 
% Finally, we fixed $F=4$, and adjust the hash table size $T$ from $2^{14}$ to $2^{19}$. We found that in general a larger the hash table lead to better reconstruction quality. However, for our problem, the differences were very small. We draw the conclusion that a good choice of these hyperparatemers depends on the data size and complexity. For a larger problem, these hyperparameters should have larger values. We repeated the same experiment on other datasets and tasks. We concluded that $T=2^{15}$, $F=4$ and $L=8$ in general give good performance for all the cases we use in this study.

\paragraph{\textbf{Hash Encoder Locations}}
In this section, we present our investigation of the impact of hash encoder positions $\mathcal{E}$ on data prediction quality. We began by studying the problem in 1D using the TSR task and subsequently move to higher dimensions using the DGS task.

% We study the impact of hash encoder positions $\mathcal{E}$ on data prediction quality. We first investigate the problem in 1D using the TSR task. Then we move to higher dimensions using the dynamic shadow task.

For the TSR task, we utilized the \textbf{vortices} dataset and construct 6 HyperINR networks with varying numbers of hash encoders, as depicted in \Cref{fig:diffgrids}AB. These hash encoders were uniformly distributed along the time axis and configured equally based on findings obtained from the experiment described in \Cref{sec:hash-encoder-param}. We then followed our knowledge distillation process, utilizing a pre-trained CoordNet to distill all the HyperINRs equally and sufficiently (\ie more than 30k epochs). Network performances were measured using PSNR and SSIM. Our results show that the knowledge of a CoordNet can be fully distilled into a HyperINR with 16 or more hash encoders. Notably, this number is close to the number of training samples in the TSR task. Further experimentation on the \textbf{pressure} and \textbf{temperature} datasets confirms that this is not a coincidence. We conclude that, for TSR tasks, the number of hash encoders needs to be comparable to the training set size to ensure good knowledge distillation performance.

Moving to higher dimensions, we conducted experiments using the DGS task with 4 different HyperINRs. For the first three networks, we used Poisson disk sampling to generate encoder positions $\mathcal{E}$ with different disk radii, resulting in approximately 100, 150, and 200 hash encoders, as shown in \Cref{fig:diffgrids}C. For the fourth model, we used approximately 200 hash encoders, but with 3/4 of them positioned using Gaussian kernels centered around training set parameters. This resulted in a stronger bias towards the training data. Our results, as shown in \Cref{fig:diffgrids}C, demonstrate that our model can significantly outperform direct data interpolation with only 100 hash encoders. But this amount of hash encoders cannot fully capture CoordNet's behavior. Nevertheless, gradually adding more hash encoders can shrink the gap between CoordNet and HyperINR, indicating that Poisson disk sampling in general can provide a good heuristic for hash encoder positions. Moreover, it is possible to further improve the model's performance and allow it to be biased towards known information. This operation can be useful if there is prior knowledge about which areas of the parameter space will be explored more.

\section{Application Performance}

% In this section, we highlight the performance of HyperINR in each applications in terms of prediction quality and inference speed. Other training related information is summarized in \Cref{tab:results}.

This section presents the performance of HyperINR in terms of prediction quality and inference speed for each application. We provide a summary of the training-related information in \Cref{tab:results}.

\begin{table}[tb]
  \caption{Training results measured using a single NVIDIA A100-40G GPU. Teacher models were trained for 30k (TSR, DGS) or 300 (NVS) epochs. Distillations were performed for 5k (TSR, DGS) or 200 (NVS) epochs. The distillation time for \textbf{temp} and \textbf{mechhand} were significantly longer because their distillation sets were larger than the GPU memory capacity, pushing a lot of data to the CPU, causing the training process to be bottlenecked by the GPU/CPU bandwidth.}
  \vspace{-0.8em}
  \label{tab:results}
  \scriptsize%
  \centering%
  \begin{tabu}{ccccccc}\toprule
    Data & Task & $T_{\text{teacher}}$ & $T_{\text{distill}}$ & $N_C$ & $N_D$ & \makecell{ Generation \\ Latency } \\
    \midrule
    Vortices & TSR & 1h04m & 19m   & 20  & 100 & 0.08ms \\
    Pressure & TSR & 1h21m & 20m   & 20  & 105 & 0.07ms \\
    Temp     & TSR & 36m   & 3h34m & 10  & 100 & 0.07ms \\
    MPAS     & NVS & 54m   & 23m   & 100 & 981 & 0.27ms \\
    MechHand & DGS & 1h47m & 4h09m & 35  & 400 & 0.28ms \\
    \bottomrule
  \end{tabu}%
\end{table}

\begin{figure*}[tb]
  \centering
  \includegraphics[width=\linewidth]{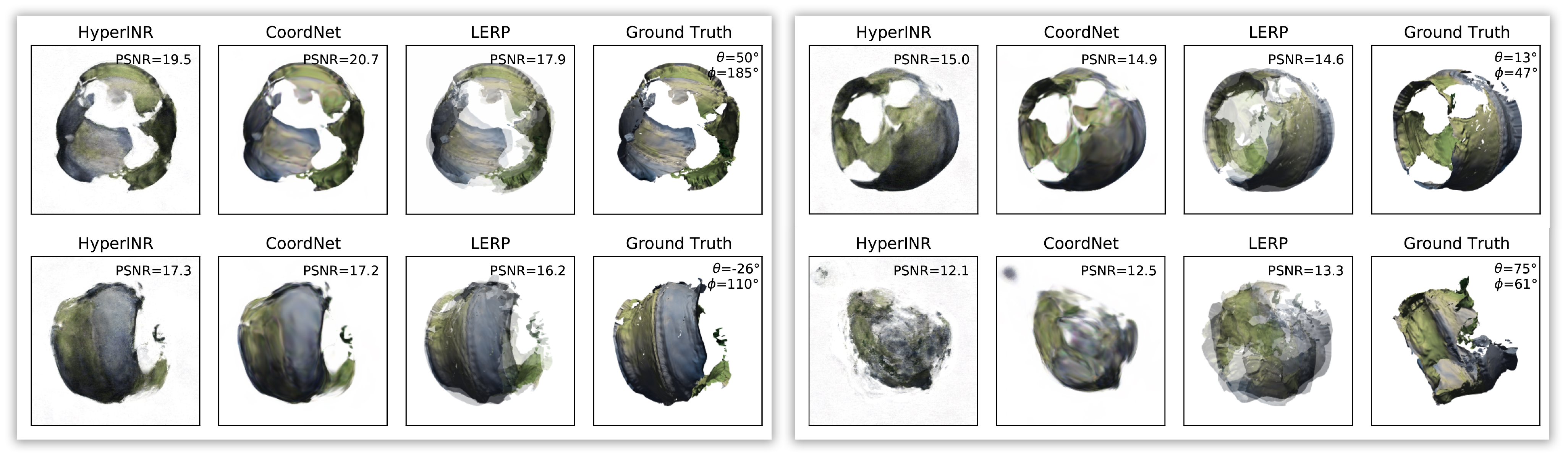}
  \vspace{-2em}
  \caption{Results for the novel view synthesis task performed on the \textbf{MPAS} dataset. The dataset was created by He~\etal~\cite{he2019insitunet}. Compared with CoordNet, HyperINR can instantaneously generate novel views while maintaining similar qualities.}
  \label{fig:mpas}
  \vspace{-1em}
\end{figure*}

\begin{figure}[tb]
  \centering
  \vspace{-0.5em}
  \includegraphics[width=0.9\linewidth]{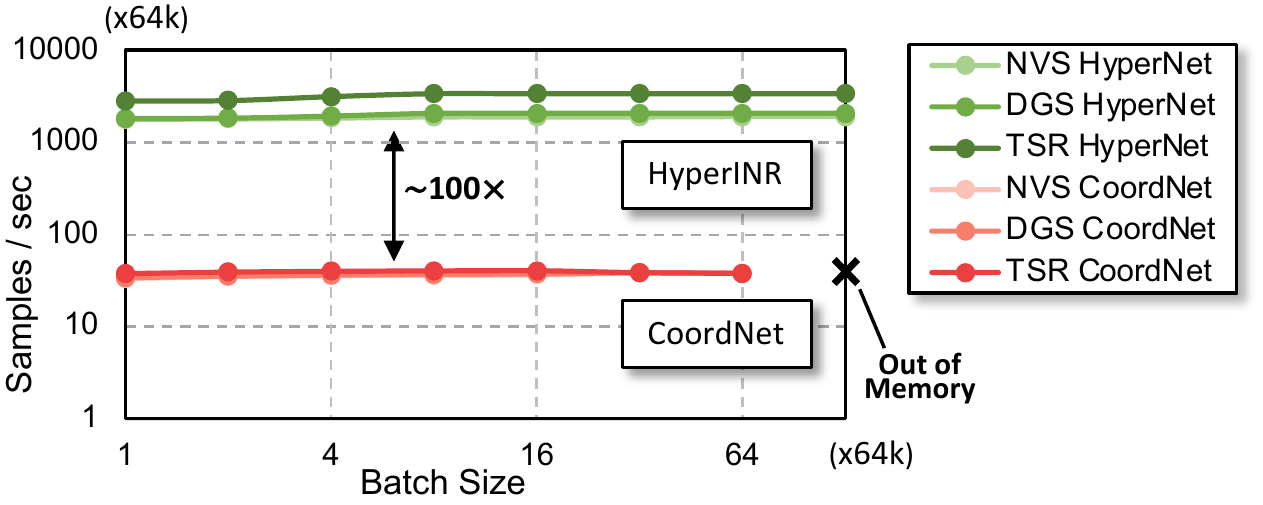}
  \vspace{-1em}
  \caption{Network inference bandwidth comparisons between CoordNet and HyperINR. Measured using a single NVIDIA A100-40G GPU.}
  \label{fig:bandwidth}
  \vspace{-0.5em}
\end{figure}

\subsection{Novel View Synthesis}

\begin{figure}[tb]
  % \hspace*{-1.5em}
  \centering
  \includegraphics[width=0.9\linewidth]{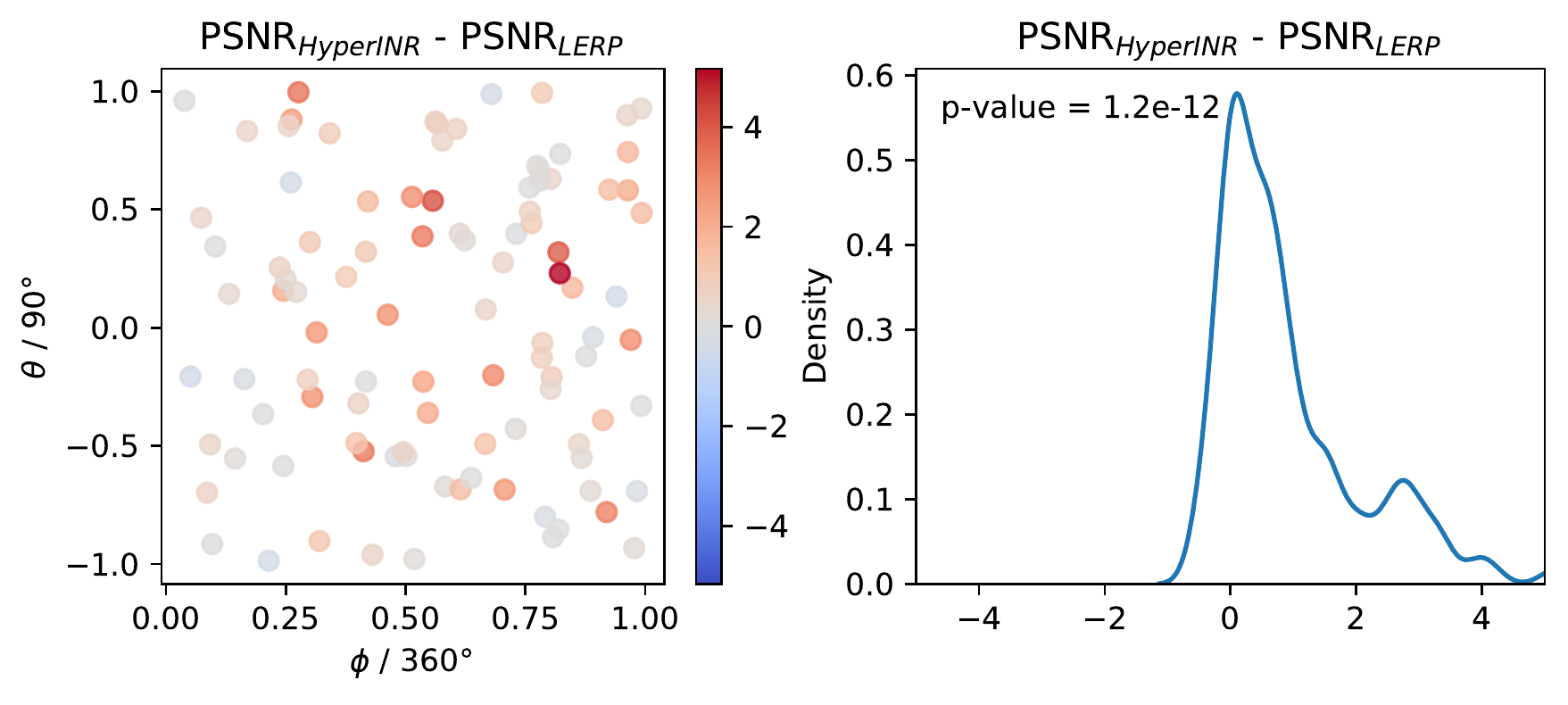}
  \vspace{-1.25em}
  \caption{Visualizations of the PSNR differences between HyperINR and LERP on the \textbf{MPAS} evaluation set. 
  Left: the scatter plot highlighting the sample location in the parameter space. We encode the PSNR differences as marker colors.
  Right: the kernel density estimation of the PSNR differences.
  Additionally, a statistical test (one-sample t-test) was performed to verify if the PSNR differences were positive.}
  \label{fig:images_vis}
\end{figure}

To evaluate the image generation quality of HyperINR, we predicted images for all parameters in the evaluation set using HyperINR and inverse distance weighting (LERP), and compared them with the actual images in the evaluation set. We then calculated the PSNR scores and highlight the differences in \Cref{fig:images_vis}. Our visualizations show that HyperINR-generated images generally have higher PSNR scores compared to LERP-generated images. We also conducted a statistical hypothesis test on whether the PSNR differences were statistically greater than zero. The p-value for the test was $1.2\times10^{-12} < 0.05$, indicating a significant difference.

The predicted images are displayed in \Cref{fig:mpas}, which also includes results generated by CoordNet for comparison.
% We display the predicted images directly in \Cref{fig:mpas} and also include predictions generated using CoordNet for comparison. 
We found that CoordNet could generally predict images at novel views well, while LERP images would produce distracting artifacts. HyperINR could avoid these artifacts but introduced some high-frequency noises. Despite these advantages, we observed that when CoordNet's performance was not satisfactory, HyperINR's performance also deteriorated, highlighting one of the limitations of our approach. Finally, we compared the network inference throughput between CoordNet and HyperINR in \Cref{fig:bandwidth} and found that our method was around $100\times$ faster.

% To investigate the image generation quality, we use HyperINR and inverse distance weighting (\ie LERP) to predict images for all parameters in the evaluation set. We compared these images with actual images in the evaluation set and computed PSNR scores.  We highlight the PSNR differences in \Cref{fig:images_vis}.  We also performed a statistical test to verify if the PSNR differences were statistically positive. Base on our visualizations, we found that images generated by HyperINR in general have higher PSNR scores. The calculated p-value for the statistical test was $1.2e^{-12} < 0.05$, indicating that the difference was statistically significant.

% We also highlight the predicted images directly in \Cref{fig:mpas}, where we also include predictions generated directly using CoordNet. We found that CoordNet in general predicted images at novel views well whereas LERP images could produce distracting artifacts. HyperINR was able to avoid these artifacts, but in introduced some high frequency noises. Despite these advantages, we also noticed that when CoordNet did not perform well, HyperINR's performance was also greatly limited. This highlights one of the limitations of our approach. 
 
% Finally, we compared the network inference throughput between CoordNet and HyperINR in \Cref{fig:bandwidth}. It indicates that our method is faster than CoordNet in terms of inference by aournd $100\times$. 

% radius = 0.060, total  177 samples

% \qwu{Show Throughput in a table.}

\subsection{Temporal Super Resolution}

\begin{figure}[tbp]
  \centering
  \includegraphics[width=0.96\linewidth]{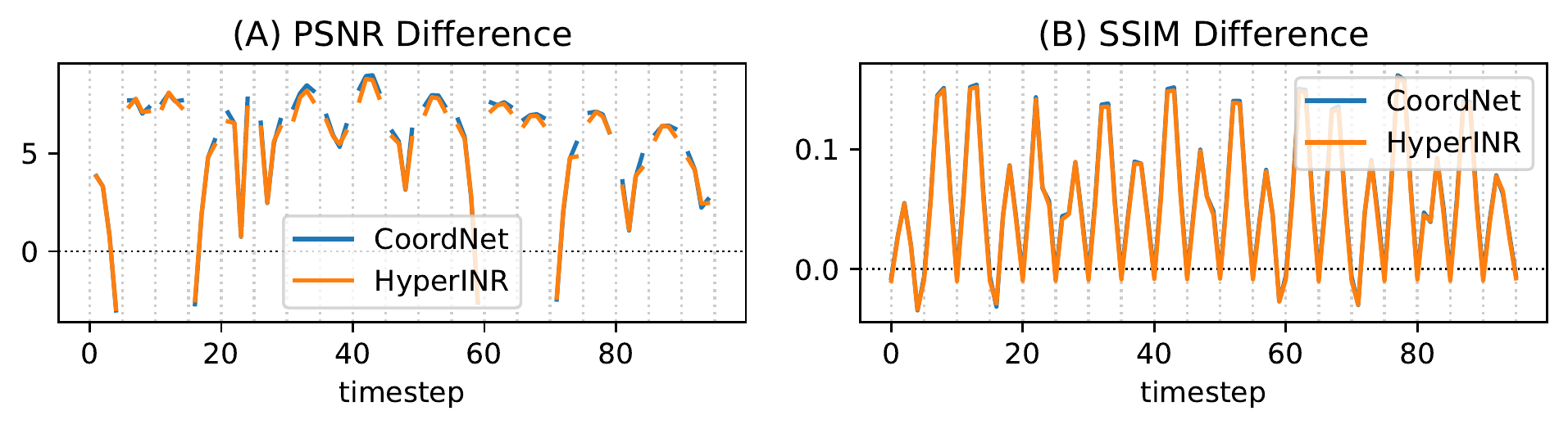}
  \vspace{-1.5em}
  \caption{Comparisons of the network reconstruction quality between HyperINR and CoordNet for the temporal super-resolution task, performed on the \textbf{vortices} dataset.}
  \vspace{-1em}
  \label{fig:timevarying_vis}
\end{figure}

\begin{figure*}[tbp]
  \centering
  \includegraphics[width=0.9\linewidth]{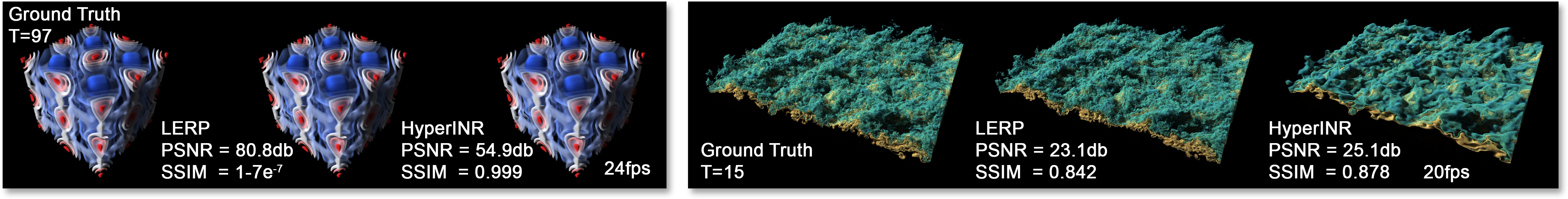}
  \vspace{-1em}
  \caption{Rendering results of  temporal super-resolution experiments using (Left) the \textbf{pressure} dataset and (Right) the \textbf{temperature} dataset. Renderings were performed on a single NVIDIA RTX8000 GPU.}
  \label{fig:tsr_others}
  \vspace{-0.5em}
\end{figure*}

We performed TSR tasks using three datasets and evaluated CoordNet, HyperINR, and LERP at all timesteps, calculating the PSNRs and SSIMs with respect to the ground truths, shown in \Cref{fig:timevarying_vis}. 
We also display volumetric path tracing  results of selected timesteps in \Cref{fig:teaser} and \Cref{fig:tsr_others} for all datasets. 

Our visualizations show that HyperINR was able to closely match the performance of CoordNet for all the datasets. Based on our volume rendering results, we found that HyperINR-generated results were generally more structurally accurate compared to LERP-generated results. However, HyperINR failed to outperform LERP on datasets that were intrinsically suitable for interpolation, such as the \textbf{pressure} dataset shown in \Cref{fig:timevarying_vis}A. For datasets containing many fine details like the \textbf{temperature} dataset shown in \Cref{fig:timevarying_vis}B, HyperINR's results were more accurate in terms of PSNR and SSIM, but high-frequency details were mostly missing on rendered images. LERP was able to generate data containing many details, but these details were inaccurate.

% We performed TSR tasks using three dataset. We evaluated CoordNet, HyperINR and LERP at all the timesteps, and calculate the PSNRs and SSIMs with respect to the ground truths. We also highlight volumetric path tracing results of selected timesteps in \Cref{fig:teaser} and \Cref{fig:tsr_others}. We found that, as shown in \Cref{fig:timevarying_vis}, our HyperINR was able to closely match the performance of CoordNet. This results were observed for all the datasets. Based on our volume rendering results, we found that, results predicted by HyperINR in general were structurally more accurate compared with LERP. However, HyperINR failed to outperform LERP on datasets that are intrinsically suitable for interpolation, such as the \textbf{pressure} dataset shown in \Cref{fig:timevarying_vis}A.  For datasets that contains many fine details like the \textbf{temp} dataset shown in \Cref{fig:timevarying_vis}B, HyperINR was able to perform better in terms of PSNR and SSIM, but high-frequency details were mostly missing on rendered images. Whereas LERP was able to generate data containing many details, but these details were inaccurate.

% In terms of performance, we found that HyperINR achieved good volume rendering performance (20-30 fps), and model inference bandwidths were also measured and presented in \Cref{fig:bandwidth}.

In terms of performance, we found that HyperINR achieved good volume rendering performance (20$\sim$30fps). Note that these results were rendered using a costly volumetric path tracing algorithm with one directional light. The same framesize (800$\times$600), samples per pixels (spp = 1) and lighting configuration (1 directional light) were used for all the experiments. Additionally, model inference bandwidths were also measured and presented in \Cref{fig:bandwidth}.

% \qwu{Connect to teaser image and \Cref{fig:timevarying_vis}}
% \qwu{Show NekRS images and related metrics + Rendering performance.}

\begin{figure*}[tbp]
  \centering
  \includegraphics[width=0.9\linewidth]{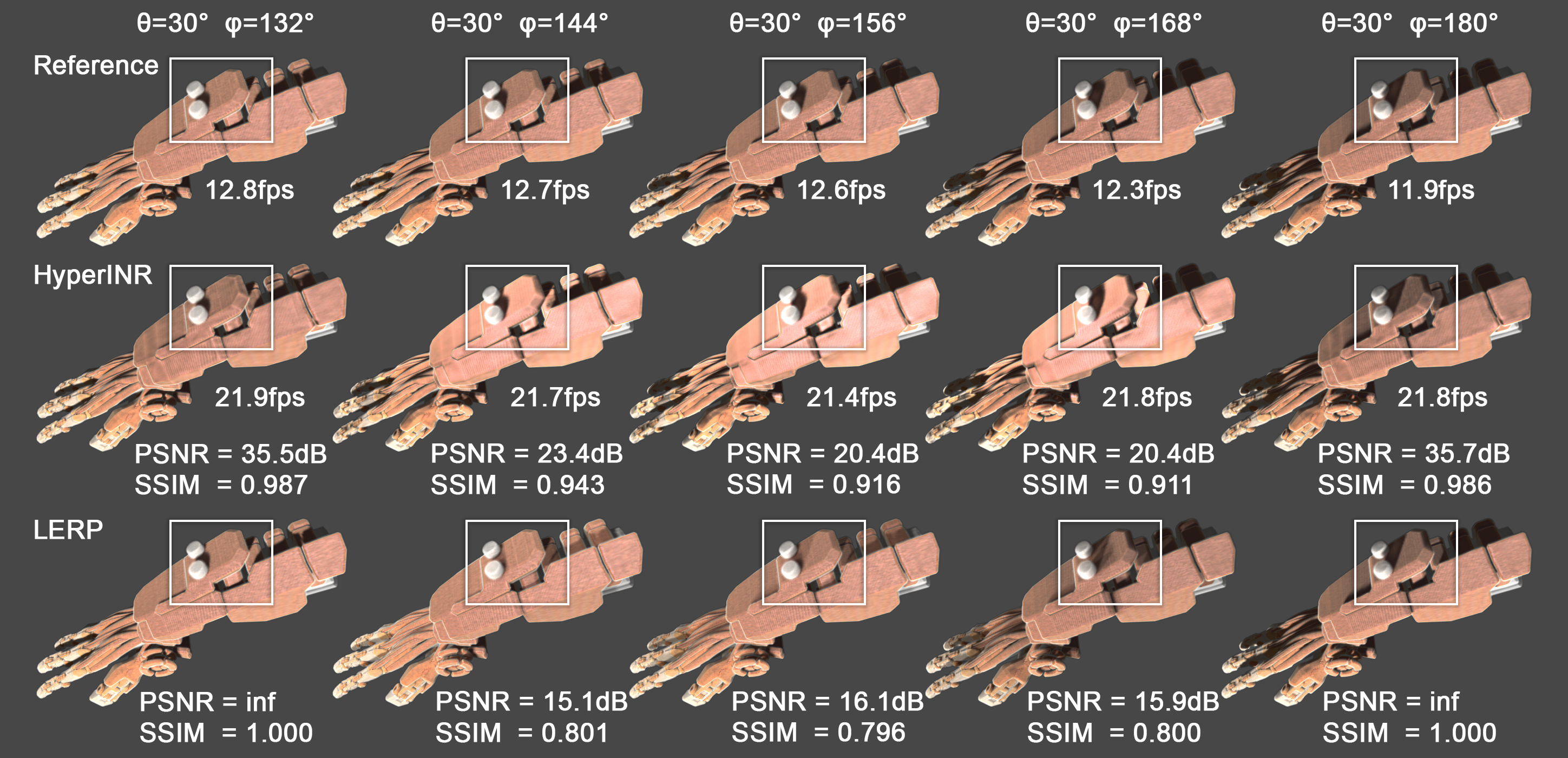}
  \vspace{-0.8em}
  \caption{Rendering results of the dynamic global shadow task. Renderings were performed on a single NVIDIA RTX8000 GPU. As highlighted by white boxes, HyperINR could more accurately predict the shadow movements, while LERP gradually faded shadows in and out.}
  \label{fig:shadowmap}
  \vspace{-0.8em}
\end{figure*}

\subsection{Dynamic Global Shadow}

For the DGS task, we used an INR to encode shadow coefficients and created global shadows for the \textbf{MechHand} dataset. We have already highlighted the PSNR and SSIM results of this experiment in \Cref{sec:ablation-params}. Here, we focus on the visual quality aspect.

% For the DGS task, we use an INR to encode shadow coefficients and create global shadows for the \textbf{MechHand} dataset. \Cref{sec:ablation-params} already highlighted the PSNR and SSIM results of this experiment. Thus we focus on the visual quality aspect in this section.

In \Cref{fig:shadowmap}, we compared global shadows generated using secondary rays (\ie the reference), HyperINR, and LERP. We gradually adjusted the azimuthal angle $\phi$ from 132$^{\circ}$ to 180$^{\circ}$ while fixing the polar angle $\theta$. The first and last data frames were presented in the training set. Our visualizations show that LERP could not correctly predict the movement of the shadow and gradually faded shadows in and out. Conversely, HyperINR was able to plausibly predict the shadow movement. However, the generated INR seemed to also make the volume brighter in some areas, perhaps due to some incorrectly predicted shadow coefficient values.

% In \Cref{fig:shadowmap} we compared global shadows generated using secondary rays (\ie the reference), HyperINR and LERP. We fixed the polar angle $\theta$, and gradually adjusted the azimuthal angle $\phi$ from 132$^{\circ}$ to 180$^{\circ}$. The first and the last data frames were presented in the training set. We found that LERP could not correctly predict the  movement of the shadow. Instead, LERP would gradually fade shadows in and out. Conversely, HyperINR was able to plausibly predict the shadow movement. However, the generated INR seemed to also make the volume brighter in some areas. This was perhaps due to some of the shadow coefficient intensities were incorrectly predicted. 

% In the same figure, we also highlight the rendering performance differences. We found that our method was able to significantly improve the rendering performance in general by avoiding secondary rays. For the tested configuration, a roughly $2\times$ speedup was observed. We also highlight the INR generation latency in \Cref{tab:results}. We can see that our method can support real-time shadow INR generations.

Additionally, we highlight the rendering performance differences in \Cref{fig:shadowmap}. We found that our method significantly improved the rendering performance by avoiding secondary rays. For the tested configuration, we observed a roughly $2\times$ speedup. Furthermore, we present the INR generation latency in \Cref{tab:results}, which indicates that our method can support real-time shadow INR generations.

\section{Discussion and Future Work}

% In this section, we summarize four key insights based on our results.
In this section, we present four key insights derived from our study.

% First, our HyperINR can achieve good  performance  in part due to the use of CoordNet, which is a very effective teacher network for data and visualization generation. However, our experiments also proved that it is possible to properly build a hypernetwork with a meaningful weight space suitable for interpolation. Such a hypernetwork can continuously approximate a high-dimensional space only using a finite  number of encoder instances. In this work, we use knowledge distillation to conduct the optimization process. However, it should be possible directly optimize a generative hypernetwork that can achieve a similar level of data generation performance. Our results makes it an even more promising research direction.

Firstly, our results generally highlight HyperINR's exceptional performance for data and visualization generation. This can be attributed to the incorporation of CoordNet, which is a highly effective teacher network. In addition, our experiments also demonstrate the possibility to construct a hypernetwork with a weight space meaningful for interpolation. Such a hypernetwork can continuously approximate a high-dimensional space utilizing a finite number of encoder instances. 
Although we employ knowledge distillation for optimization in this study, directly optimizing a generative hypernetwork to achieve comparable data generation performance should be feasible and represents a promising research direction.

% 
% While we employ knowledge distillation for optimization in this study, directly optimizing a generative hypernetwork to achieve comparable data generation performance should be also possible. This can be a promising research direction.
% 

% Secondly, our HyperINR uses a large number of small multiresolution hash encoders to approximate a high-dimensional space, which is more flexible and effective than using a single large encoder. This finding can be supported by our ablation study (\Cref{sec:ablation-canilla}). This property makes HyperINR suitable for learning an ensumble of data that comprises many different data frames.

% Second, our HyperINR employs many small multiresolution hash encoders to approximate data in a high-dimensional  space, which is more flexible flexible and effective than utilizing a single large encoder. This observation is supported by our ablation study (refer to \Cref{sec:ablation-canilla}). This characteristic renders HyperINR suitable for learning ensembles of data consisting of various data frames.
% However, as emphasized in \Cref{sec:ablation-params}, an efficient construction of encoder positions can be tricky to achieve, and often requires a lot of experiments. In high dimensions, having more encoders does not linearly makes HyperINR perform better. A more general and effective approach to compute encoder positions is much needed.

Secondly, our HyperINR leverages numerous small multiresolution hash encoders to approximate data modeled by high-dimensional parameters, providing a more flexible and effective approach compared to using a single large encoder. This observation is corroborated by our ablation study, as detailed in \Cref{sec:ablation-canilla}. This attribute makes HyperINR well-suited for learning ensembles of data comprising diverse data frames.
Nonetheless, as highlighted in \Cref{sec:ablation-params}, efficiently constructing encoder positions can be challenging and often necessitates extensive experimentation. In higher dimensions, the addition of  encoders does not always result in a linear improvement in HyperINR's performance. Therefore, a more general and effective method for computing encoder positions is highly desirable.

% Third, as shown in \Cref{tab:results}, the knowledge distillation to train a HyperINR can be time consuming due to two reasons. First, to achieve good distillation quality, a very large distillation set might be needed, prolonging the training process. Second, if the distillation set is generated on-demand, the training speed can be greatly limited by the inference bandwidth of CoordNet. We circumvented this bottleneck by pre-computing the distillation set, bring up to $8\times$ speedup on an A100 for some cases. But this method is still inefficient when  the distillation set size exceeds the GPU memory capacity, pushing training data to CPU, causing the training process being bottlenecked by the CPU/GPU bandwidth.
% Future studies on a more effective distillation set generation method would be interesting.

Thirdly, as demonstrated in \Cref{tab:results}, knowledge distillation employed to train HyperINR can be time-consuming due to two factors. For one, achieving high distillation quality may require a large distillation set, prolonging the training process. Secondly, when the distillation set is generated on-demand, the training speed can be significantly constrained by the inference bandwidth of CoordNet. We alleviated this bottleneck by pre-computing the distillation set, resulting in up to an $8\times$ speedup on an NVIDIA A100 for certain cases. However, this optimization remains inefficient when the distillation set size surpasses the GPU memory capacity, pushing training data to the CPU and causing the training process to be constricted by the CPU/GPU bandwidth. Exploring more effective distillation set generation methods warrants further investigation.

% Finally, although knowledge distillation in general can greatly improve HyperINR's data generation capability, it also makes HyperINR's performance upper-limited by the teacher network. 

Lastly, while knowledge distillation can substantially enhance HyperINR's data generation capabilities, it also imposes an upper limit on its performance.
% In situations where the teacher network underperforms, the performance of HyperINR can also be significantly impeded, as shown in \Cref{fig:tsr_others}.
In cases where the teacher network underperforms, HyperINR's performance may also be considerably hindered, as illustrated in \Cref{fig:tsr_others}. 
% Exploring more flexible training strategies constitutes a valuable direction for future research.
Investigating more flexible training strategies represents a promising avenue for future research.

% Secondly, directly generalizing a single grid-based parametric encoder to higher dimensions often requires exponentially increase the number of parameters, because the number of encoding parameters should be somewhat proportional to the input space size. Our method breaks this limitation by virtually positioning encoders as a ``point-cloud'' in high-dimension, making it more memory efficient.

\section{Conclusion}

% In this work, we introduce HyperINR, which is a hypernetwork that can directly predict the weight of a regular INR for unseen scene parameters. By utilizing a set of small multiresolution hash encoders, a shared MLP, and a deeply embedded weight interpolation operation, HyperINR can achieve the $100\times$ faster inference bandwidth. The predicted INR can also be volume rendered interactively with global illumination. Furthermore, HyperINR can also achieve the state-of-the-art data generation performance via knowledge distillation. 
% Our approach demonstrates promising potential for further application in a wide range of scientific visualization contexts.

% To be concluded, our work presents HyperINR, a novel hypernetwork that enables conditional generations of INRs for unseen scene parameters. This is made possible through the utilization of many small multiresolution hash encoders, a shared MLP, and a deeply embedded weight interpolation operation, resulting in an impressive $100\times$ higher inference bandwidth and the support of  interactive volume rendering with high realism. Additionally, our approach also achieves state-of-the-art data generation performance via knowledge distillation. Finally, our findings highlight the potential of HyperINR for a variety of scientific visualization tasks, demonstrating its effectiveness and efficiency. We believe that HyperINR represents a  step forward in the development of INR-based approaches for scientific visualization and beyond.

We introduce HyperINR, an innovative hypernetwork facilitating conditional generation of INRs for unseen scene parameters. Enabled by the employment of numerous small multiresolution hash encoders, a shared MLP, and a deeply embedded weight interpolation operation, HyperINR achieves an impressive $100\times$ higher inference bandwidth and  interactive volume rendering with exceptional realism. Moreover, our method attains state-of-the-art data and visualization generation performance through knowledge distillation. Our results underscore the potential of HyperINR in various visualization tasks, showcasing its effectiveness and efficiency. We believe that HyperINR represents a step forward in the development of implicit neural representation based approaches for the field of scientific visualization and beyond.

\acknowledgments{%
This research was supported in part by the Department of Energy through grant 
DE-SC0019486 and an Intel oneAPI Centers of Excellence grant.
The authors also express sincere gratitudes to Weishen Liu (UC Davis Alumni) and 
Daniel Zavorotny (UC Davis) for their assistance with data preparation.%
}

\bibliographystyle{abbrv-doi-hyperref}

\bibliography{main}

\begin{thebibliography}{10}

\bibitem{barron2021mip}
J.~T. Barron, B.~Mildenhall, M.~Tancik, P.~Hedman, R.~Martin-Brualla, and P.~P.
  Srinivasan.
\newblock {Mip-NeRF}: A multiscale representation for anti-aliasing neural
  radiance fields.
\newblock In {\em Proceedings of the IEEE/CVF International Conference on
  Computer Vision}, pp. 5855--5864, 2021.

\bibitem{bauer2023fovolnet}
D.~Bauer, Q.~Wu, and K.-L. Ma.
\newblock Fovolnet: Fast volume rendering using foveated deep neural networks.
\newblock {\em IEEE Transactions on Visualization and Computer Graphics},
  29(1):515--525, 2023. \href{https://doi.org/10.1109/TVCG.2022.3209498}
{doi: {{%
10\hspace{.1pt}\discretionary{.}{%
}{.}\hspace{.4pt}1109\discretionary{/}{%
}{/}TVCG\hspace{.1pt}\discretionary{.}{%
}{.}\hspace{.4pt}2022\hspace{.1pt}\discretionary{.}{%
}{.}\hspace{.4pt}3209498}}}


\bibitem{berger2018generative}
M.~Berger, J.~Li, and J.~A. Levine.
\newblock A generative model for volume rendering.
\newblock {\em IEEE transactions on visualization and computer graphics},
  25(4):1636--1650, 2018.

\bibitem{bertinetto2016learning}
L.~Bertinetto, J.~F. Henriques, J.~Valmadre, P.~Torr, and A.~Vedaldi.
\newblock Learning feed-forward one-shot learners.
\newblock {\em Advances in neural information processing systems}, 29, 2016.

\bibitem{bridson2007fast}
R.~Bridson.
\newblock Fast poisson disk sampling in arbitrary dimensions.
\newblock {\em SIGGRAPH sketches}, 10(1):1, 2007.

\bibitem{bucilua2006model}
C.~Bucilu{\u{a}}, R.~Caruana, and A.~Niculescu-Mizil.
\newblock Model compression.
\newblock In {\em Proceedings of the 12th ACM SIGKDD international conference
  on Knowledge discovery and data mining}, pp. 535--541, 2006.

\bibitem{engel2020deep}
D.~Engel and T.~Ropinski.
\newblock Deep volumetric ambient occlusion.
\newblock {\em IEEE Transactions on Visualization and Computer Graphics},
  27(2):1268--1278, 2020.

\bibitem{gehring2017convolutional}
J.~Gehring, M.~Auli, D.~Grangier, D.~Yarats, and Y.~N. Dauphin.
\newblock Convolutional sequence to sequence learning.
\newblock In {\em International Conference on Machine Learning}, pp.
  1243--1252. PMLR, 2017.

\bibitem{guo2020ssr}
L.~Guo, S.~Ye, J.~Han, H.~Zheng, H.~Gao, D.~Z. Chen, J.-X. Wang, and C.~Wang.
\newblock {SSR-VFD}: Spatial super-resolution for vector field data analysis
  and visualization.
\newblock In {\em Proceedings of IEEE Pacific Visualization Symposium}, 2020.

\bibitem{ha2016hypernetworks}
D.~Ha, A.~Dai, and Q.~V. Le.
\newblock Hypernetworks.
\newblock {\em arXiv preprint arXiv:1609.09106}, 2016.

\bibitem{hadadan2021neural}
S.~Hadadan, S.~Chen, and M.~Zwicker.
\newblock Neural radiosity.
\newblock {\em ACM Transactions on Graphics (TOG)}, 40(6):1--11, 2021.

\bibitem{han2019tsr}
J.~Han and C.~Wang.
\newblock {TSR-TVD}: Temporal super-resolution for time-varying data analysis
  and visualization.
\newblock {\em IEEE transactions on visualization and computer graphics},
  26(1):205--215, 2019.

\bibitem{han2020ssr}
J.~Han and C.~Wang.
\newblock {SSR-TVD}: Spatial super-resolution for time-varying data analysis
  and visualization.
\newblock {\em IEEE Transactions on Visualization and Computer Graphics}, 2020.

\bibitem{han2022coordnet}
J.~Han and C.~Wang.
\newblock Coordnet: Data generation and visualization generation for
  time-varying volumes via a coordinate-based neural network.
\newblock {\em IEEE Transactions on Visualization and Computer Graphics}, 2022.

\bibitem{han2022vcnet}
J.~Han and C.~Wang.
\newblock Vcnet: A generative model for volume completion.
\newblock {\em Visual Informatics}, 6(2):62--73, 2022.

\bibitem{han2021stnet}
J.~Han, H.~Zheng, D.~Z. Chen, and C.~Wang.
\newblock {STNet}: An end-to-end generative framework for synthesizing
  spatiotemporal super-resolution volumes.
\newblock {\em IEEE Transactions on Visualization and Computer Graphics},
  28(1):270--280, 2021.

\bibitem{he2019insitunet}
W.~He, J.~Wang, H.~Guo, K.-C. Wang, H.-W. Shen, M.~Raj, Y.~S. Nashed, and
  T.~Peterka.
\newblock Insitunet: Deep image synthesis for parameter space exploration of
  ensemble simulations.
\newblock {\em IEEE transactions on visualization and computer graphics},
  26(1):23--33, 2019.

\bibitem{hinton2015distilling}
G.~Hinton, O.~Vinyals, and J.~Dean.
\newblock Distilling the knowledge in a neural network.
\newblock {\em arXiv preprint arXiv:1503.02531}, 2015.

\bibitem{jain2017compressed}
S.~Jain, W.~Griffin, A.~Godil, J.~W. Bullard, J.~Terrill, and A.~Varshney.
\newblock Compressed volume rendering using deep learning.
\newblock In {\em Proceedings of the Large Scale Data Analysis and
  Visualization (LDAV) Symposium. Phoenix, AZ}, 2017.

\bibitem{kim2022neuralvdb}
D.~Kim, M.~Lee, and K.~Museth.
\newblock Neuralvdb: High-resolution sparse volume representation using
  hierarchical neural networks.
\newblock {\em arXiv preprint arXiv:2208.04448}, 2022.

\bibitem{klocek2019hypernetwork}
S.~Klocek, {\L}.~Maziarka, M.~Wo{\l}czyk, J.~Tabor, J.~Nowak, and
  M.~{\'S}mieja.
\newblock Hypernetwork functional image representation.
\newblock In {\em Artificial Neural Networks and Machine Learning--ICANN 2019:
  Workshop and Special Sessions: 28th International Conference on Artificial
  Neural Networks, Munich, Germany, September 17--19, 2019, Proceedings 28},
  pp. 496--510. Springer, 2019.

\bibitem{littwin2019deep}
G.~Littwin and L.~Wolf.
\newblock Deep meta functionals for shape representation.
\newblock In {\em Proceedings of the IEEE/CVF International Conference on
  Computer Vision}, pp. 1824--1833, 2019.

\bibitem{lu2021compressive}
Y.~Lu, K.~Jiang, J.~A. Levine, and M.~Berger.
\newblock Compressive neural representations of volumetric scalar fields.
\newblock vol.~40, pp. 135--146, 2021. \href{https://doi.org/10.1111/cgf.14295}
{doi: {{%
10\hspace{.1pt}\discretionary{.}{%
}{.}\hspace{.4pt}1111\discretionary{/}{%
}{/}cgf\hspace{.1pt}\discretionary{.}{%
}{.}\hspace{.4pt}14295}}}


\bibitem{martel2021acorn}
J.~N. Martel, D.~B. Lindell, C.~Z. Lin, E.~R. Chan, M.~Monteiro, and
  G.~Wetzstein.
\newblock Acorn: Adaptive coordinate networks for neural scene representation.
\newblock {\em arXiv preprint arXiv:2105.02788}, 2021.

\bibitem{micikevicius2017mixed}
P.~Micikevicius, S.~Narang, J.~Alben, G.~Diamos, E.~Elsen, D.~Garcia,
  B.~Ginsburg, M.~Houston, O.~Kuchaiev, G.~Venkatesh, et~al.
\newblock Mixed precision training.
\newblock {\em arXiv preprint arXiv:1710.03740}, 2017.

\bibitem{mildenhall2020nerf}
B.~Mildenhall, P.~P. Srinivasan, M.~Tancik, J.~T. Barron, R.~Ramamoorthi, and
  R.~Ng.
\newblock Nerf: Representing scenes as neural radiance fields for view
  synthesis.
\newblock In {\em European conference on computer vision}, pp. 405--421.
  Springer, 2020.

\bibitem{tiny-cuda-nn}
T.~M\"uller.
\newblock Tiny {CUDA} neural network framework, 2021 (Online).
\newblock https://github.com/nvlabs/tiny-cuda-nn.

\bibitem{muller2022instant}
T.~M{\"u}ller, A.~Evans, C.~Schied, and A.~Keller.
\newblock Instant neural graphics primitives with a multiresolution hash
  encoding.
\newblock {\em ACM Transactions on Graphics (ToG)}, 41(4):1--15, 2022.

\bibitem{muller2019neural}
T.~M{\"u}ller, B.~McWilliams, F.~Rousselle, M.~Gross, and J.~Nov{\'a}k.
\newblock Neural importance sampling.
\newblock {\em ACM Transactions on Graphics (TOG)}, 38(5):1--19, 2019.

\bibitem{muller2021real}
T.~M{\"u}ller, F.~Rousselle, J.~Nov{\'a}k, and A.~Keller.
\newblock Real-time neural radiance caching for path tracing.
\newblock {\em arXiv preprint arXiv:2106.12372}, 2021.

\bibitem{nguyen2020hypervae}
P.~Nguyen, T.~Tran, S.~Gupta, S.~Rana, H.-C. Dam, and S.~Venkatesh.
\newblock Hypervae: A minimum description length variational hyper-encoding
  network.
\newblock {\em CoRR}, 2020.

\bibitem{oh2020hcnaf}
G.~Oh and J.-S. Valois.
\newblock Hcnaf: Hyper-conditioned neural autoregressive flow and its
  application for probabilistic occupancy map forecasting.
\newblock In {\em Proceedings of the IEEE/CVF Conference on Computer Vision and
  Pattern Recognition}, pp. 14550--14559, 2020.

\bibitem{ratzlaff2019hypergan}
N.~Ratzlaff and L.~Fuxin.
\newblock Hypergan: A generative model for diverse, performant neural networks.
\newblock In {\em International Conference on Machine Learning}, pp.
  5361--5369. PMLR, 2019.

\bibitem{s3d}
M.~Rieth, A.~Gruber, F.~A. Williams, and J.~H. Chen.
\newblock Enhanced burning rates in hydrogen-enriched turbulent premixed flames
  by diffusion of molecular and atomic hydrogen.
\newblock {\em Combustion and Flame}, p. 111740, 2021.

\bibitem{shepard1968two}
D.~Shepard.
\newblock A two-dimensional interpolation function for irregularly-spaced data.
\newblock In {\em Proceedings of the 1968 23rd ACM national conference}, pp.
  517--524, 1968.

\bibitem{shi2022vdl}
N.~Shi, J.~Xu, H.~Li, H.~Guo, J.~Woodring, and H.-W. Shen.
\newblock Vdl-surrogate: A view-dependent latent-based model for parameter
  space exploration of ensemble simulations.
\newblock {\em IEEE Transactions on Visualization and Computer Graphics},
  29(1):820--830, 2022.

\bibitem{sitzmann2019siren}
V.~Sitzmann, J.~N. Martel, A.~W. Bergman, D.~B. Lindell, and G.~Wetzstein.
\newblock Implicit neural representations with periodic activation functions.
\newblock In {\em arXiv}, 2020.

\bibitem{sitzmann2019scene}
V.~Sitzmann, M.~Zollh{\"o}fer, and G.~Wetzstein.
\newblock Scene representation networks: Continuous 3d-structure-aware neural
  scene representations.
\newblock {\em Advances in Neural Information Processing Systems}, 32, 2019.

\bibitem{skorokhodov2021adversarial}
I.~Skorokhodov, S.~Ignatyev, and M.~Elhoseiny.
\newblock Adversarial generation of continuous images.
\newblock In {\em Proceedings of the IEEE/CVF Conference on Computer Vision and
  Pattern Recognition}, pp. 10753--10764, 2021.

\bibitem{takikawa2021neural}
T.~Takikawa, J.~Litalien, K.~Yin, K.~Kreis, C.~Loop, D.~Nowrouzezahrai,
  A.~Jacobson, M.~McGuire, and S.~Fidler.
\newblock Neural geometric level of detail: Real-time rendering with implicit
  3d shapes.
\newblock In {\em Proceedings of the IEEE/CVF Conference on Computer Vision and
  Pattern Recognition}, pp. 11358--11367, 2021.

\bibitem{tancik2020fourier}
M.~Tancik, P.~Srinivasan, B.~Mildenhall, S.~Fridovich-Keil, N.~Raghavan,
  U.~Singhal, R.~Ramamoorthi, J.~Barron, and R.~Ng.
\newblock Fourier features let networks learn high frequency functions in low
  dimensional domains.
\newblock {\em Advances in Neural Information Processing Systems},
  33:7537--7547, 2020.

\bibitem{tikhonova2010explorable}
A.~Tikhonova, C.~D. Correa, and K.-L. Ma.
\newblock Explorable images for visualizing volume data.
\newblock {\em PacificVis}, 10:177--184, 2010.

\bibitem{vaswani2017attention}
A.~Vaswani, N.~Shazeer, N.~Parmar, J.~Uszkoreit, L.~Jones, A.~N. Gomez,
  {\L}.~Kaiser, and I.~Polosukhin.
\newblock Attention is all you need.
\newblock {\em Advances in neural information processing systems}, 30, 2017.

\bibitem{von2019continual}
J.~Von~Oswald, C.~Henning, J.~Sacramento, and B.~F. Grewe.
\newblock Continual learning with hypernetworks.
\newblock {\em arXiv preprint arXiv:1906.00695}, 2019.

\bibitem{wang2022dl4scivis}
C.~Wang and J.~Han.
\newblock Dl4scivis: A state-of-the-art survey on deep learning for scientific
  visualization.
\newblock {\em IEEE Transactions on Visualization and Computer Graphics}, 2022.

\bibitem{weiss2019volumetric}
S.~Weiss, M.~Chu, N.~Thuerey, and R.~Westermann.
\newblock Volumetric isosurface rendering with deep learning-based
  super-resolution.
\newblock {\em IEEE transactions on visualization and computer graphics},
  27(6):3064--3078, 2019.

\bibitem{weiss2021fast}
S.~Weiss, P.~Hermüller, and R.~Westermann.
\newblock Fast neural representations for direct volume rendering.
\newblock {\em Computer Graphics Forum}, 41(6):196--211, 2022.
  \href{https://doi.org/10.1111/cgf.14578}
{doi: {{%
10\hspace{.1pt}\discretionary{.}{%
}{.}\hspace{.4pt}1111\discretionary{/}{%
}{/}cgf\hspace{.1pt}\discretionary{.}{%
}{.}\hspace{.4pt}14578}}}


\bibitem{wu2022instant}
Q.~Wu, D.~Bauer, M.~J. Doyle, and K.-L. Ma.
\newblock Instant neural representation for interactive volume rendering.
\newblock {\em arXiv preprint arXiv:2207.11620}, 2022.

\bibitem{wurster2021deep}
S.~W. Wurster, H.-W. Shen, H.~Guo, T.~Peterka, M.~Raj, and J.~Xu.
\newblock Deep hierarchical super-resolution for scientific data reduction and
  visualization.
\newblock {\em arXiv preprint arXiv:2107.00462}, 2021.

\bibitem{zhang2018graph}
C.~Zhang, M.~Ren, and R.~Urtasun.
\newblock Graph hypernetworks for neural architecture search.
\newblock {\em arXiv preprint arXiv:1810.05749}, 2018.

\bibitem{zhou2017volume}
Z.~Zhou, Y.~Hou, Q.~Wang, G.~Chen, J.~Lu, Y.~Tao, and H.~Lin.
\newblock Volume upscaling with convolutional neural networks.
\newblock In {\em Proceedings of the Computer Graphics International
  Conference}, pp. 1--6, 2017.

\end{thebibliography}

%% ^^^^^   FOR IEEE VIS, EVERYTHING HERE MAY BE INCLUDED IN THE    ^^^^^ %%
%% 2-PAGE ALLOTMENT FOR REFERENCES, FIGURE CREDITS, AND ACKNOWLEDGEMENTS %%

% \appendix % You can use the `hideappendix` class option to skip everything after \appendix

% \section{About Appendices}
% Refer to \cref{sec:appendices_inst} for instructions regarding appendices.

\end{document}